\documentclass[12pt,a4paper]{article}
\usepackage{amsmath,amssymb,amsthm}
\usepackage[margin=1.0in]{geometry}
\usepackage{cite}
\usepackage{graphicx}
\usepackage{enumerate}
\allowdisplaybreaks
\usepackage[colorlinks=true
,urlcolor=blue
,anchorcolor=blue
,citecolor=blue
,filecolor=blue
,linkcolor=blue
,menucolor=blue
,pagecolor=blue
,linktocpage=true
,pdfproducer=medialab
,pdfa=true
]{hyperref}
\numberwithin{equation}{section}

\def\a{\alpha}

\def\d{\delta}

\def\l{\lambda}

\def\r{\rho}

\def\be{\begin{equation}}
\def\ee{\end{equation}}
\def\bea{\begin{eqnarray}}
\def\eea{\end{eqnarray}}

\def\lp{\left(}
\def\rp{\right)}

\def\nn{\nonumber}

\def\eg{{\it e.g., }}

\makeatletter
\renewcommand\section{\@startsection {section}{1}{\z@}%
	{-3.5ex \@plus -1ex \@minus -.2ex}
	{2.3ex \@plus.2ex}%
	{\normalfont\large\bfseries}}
\renewcommand\subsection{\@startsection{subsection}{2}{\z@}%
	{-3.25ex\@plus -1ex \@minus -.2ex}%
	{1.5ex \@plus .2ex}%
	{\normalfont\bfseries}}
\makeatother

\begin{document}

\begin{center}
\addtolength{\baselineskip}{.5mm}
\thispagestyle{empty}
\begin{flushright}
\end{flushright}

\vspace{20mm}

{\Large \bf Wormhole geometries in Einstein-aether theory}
\\[15mm]
Hanif Golchin,${}^{a,}$\footnote{h.golchin@uk.ac.ir} {Hamid R. Bakhtiarizadeh,${}^{b,}$\footnote{h.bakhtiarizadeh@kgut.ac.ir}} and Mohammad Reza Mehdizadeh${}^{a,}$\footnote{mehdizadeh.mr@uk.ac.ir}
\\[5mm]
{\it ${}^a$ Department of Physics, Faculty of Science, Shahid Bahonar University of Kerman, P.O. Box 76175, Kerman, Iran\\ ${}^b$Department of Nanotechnology, Graduate University of Advanced Technology,\\ Kerman, Iran}

\vspace{20mm}

{\bf  Abstract}
\end{center}

We present the first analysis of traversable wormholes within the framework of Einstein–aether theory. We show that the corresponding field equations admit three distinct wormhole geometries, obtained by adopting three different classes of combinations for the aether coupling constants. We examine the null and weak energy conditions for three types of wormhole shape functions. 
Our results show that, for appropriate choices of the wormhole parameters and the coupling constants of the theory, these wormhole geometries can satisfy null and weak energy conditions. Furthermore, we find that the satisfaction of these energy conditions is restricted to certain ranges of the aether coupling constants, which can be interpreted as constraints arising from the requirement of avoiding exotic matter. However, these ranges are incompatible with the constraints imposed by physical considerations and observational bounds. This suggests that, as in the Einstein gravity, exotic matter is necessary to support traversable wormhole geometries in Einstein–aether theory.

 
\vfill
\newpage


\section{Introduction}\label{int}

A wormhole is a hypothetical topological feature of spacetime that creates shortcuts between two distinct spacetimes or between two distant points within a single spacetime. The first solution describing such a structure was found by Flamm in 1916 in the context of General Relativity (GR) \cite{Flamm:2015ogy}. In 1935, Einstein and Rosen constructed an unstable, non-traversable wormhole, known as the Einstein-Rosen bridge \cite{Einstein:1935tc}. They discovered a tunnel-like geometry through which a black hole could form a bridge to a remote region of spacetime. The term ``wormhole" was first introduced in 1957 in the seminal papers of Misner and Wheeler \cite{Misner:1957mt,Misner:1960zz} to describe a mechanism for ``charge without charge." They analyzed the Riemannian geometry of manifolds with nontrivial topology.

The study of Lorentzian wormholes in the context of GR dates back to the foundational work of Morris and Thorne in 1988. Motivated by the possibility of rapid interstellar travel \cite{Morris:1988cz,Morris:1988tu}, and by introducing a static, spherically symmetric line element, they showed that exact solutions representing wormhole geometries could be found by solving the Einstein field equations. It was also found that these traversable wormholes possess a stress-energy tensor that violates the Null Energy Condition (NEC) and thus, traversable wormholes are only possible if so-called ``exotic matter" exists at their throat, involving an energy-momentum tensor that violates the NEC (see, \eg \cite{Visser:1995cc} or \cite{Lobo:2007zb} for more recent reviews). From a theoretical standpoint, quantum field theory allows for the existence of exotic energy, as demonstrated by the Casimir effect \cite{Casimir:1948dh}. Wormhole solutions supported by Casimir energy have been extensively studied in Ref. \cite{Garattini:2023qyo}. Exotic matter also appears in cosmological scenarios. For instance, phantom energy exhibits the exotic property of supporting wormhole geometries \cite{Lobo:2005yv,Lobo:2005us,Kuhfittig:2006xj,Sushkov:2005kj}. For static wormholes in Einstein gravity, the null energy condition is violated; therefore, several attempts have been made to overcome this issue.

One of the most significant challenges in wormhole physics concerns the fulfillment of the standard energy conditions. Consequently, many studies have sought realistic matter sources that can support wormhole configurations or minimize the need for exotic matter. Research in this area has focused on the construction of thin-shell, dynamical, and rotating wormholes \cite{Garcia:2011aa,Richarte:2007zz,Teo:1998dp,Kashargin:2008pk,Kar:1995ss,Arellano:2006ex,Sushkov:2007me}. However, in this work, we focus on modified gravity theories, in which the effective framework allows for static, spherically symmetric, and traversable wormhole solutions while reducing their dependence on exotic matter. Higher-order curvature terms have also been shown to allow the construction of thin-shell wormholes supported by ordinary matter \cite{Mazharimousavi:2010bm,Mazharimousavi:2010bf,Mehdizadeh:2015dta}. Similar studies have explored wormhole geometries in Brans-Dicke theory \cite{Lobo:2010sb,Sushkov:2011zh}, $ f(R) $ gravity \cite{Godani:2018blx,Pavlovic:2014gba,Eiroa:2016zjx}, Kaluza-Klein gravity \cite{Dzhunushaliev:1998ya,deLeon:2009pu}, Rastall gravity \cite{Moradpour:2016ubd}, scalar-tensor gravity \cite{Shaikh:2016dpl,Chew:2018vjp,Bahamonde:2016jqq}, solutions in the presence of a cosmological constant \cite{Anabalon:2012tu,Lemos:2003jb}, non-commutative geometry \cite{Rahaman:2013ywa,Zubair:2017hsq} and other modified gravity theories \cite{Moraes:2017mir,Sahoo:2017ual,Moradpour:2016ubd,Zubair:2016cde}.

In the present work, we aim to consider the Einstein-Aether (EA) theory and investigate the effects of it on the possibility of having wormhole solutions with normal matter. In the following, we briefly review the EA theory and its black hole solutions.

The Einstein–Aether theory (commonly referred to as æ-theory) is a generally covariant modification of general relativity that introduces a dynamical, unit timelike vector field, the “æther”, which establishes a preferred reference frame and thus violates Lorentz boost invariance. This theory was first clearly formulated by Ted Jacobson and collaborators to explore Lorentz-violating effects in the gravitational sector while preserving diffeomorphism invariance \cite{Jacobson:2004ts,Eling:2004dk}.

Research shows the existence of static, spherically symmetric black hole solutions in EA theory. These solutions closely resemble the Schwarzschild geometry outside the metric horizon, with deviations typically at the percent level for parameter regions consistent with empirical bounds. Notably, these black holes possess a distinctive universal horizon, a surface deeper than the metric horizon that traps even modes propagating arbitrarily fast relative to the æther \cite{Barausse:2011pu,Wang:2022yvi}.

Early studies by Eling and Jacobson revealed that regular, asymptotically flat black hole families exist, provided regularity at both the metric and spin-$ 0 $ horizons; numerical solutions show oscillatory æther behavior near singularities \cite{Eling:2006ec}. Further refinements considering parameterized post-Newtonian constraints demonstrate scenarios where spin-$ 0 $ horizons lie inside the metric horizon, allowing significant deviations in ADM mass, innermost stable circular orbit, and Hawking temperature-though such features may remain observationally elusive \cite{Tamaki:2007kz}. Dynamic scenarios, including gravitational collapse, have also been numerically simulated by Garfinkle, Eling, and Jacobson, confirming that black holes form with regular æther configurations under typical conditions \cite{Garfinkle:2007bk}.

The ringdown properties of black holes in Einstein–æther theory have been investigated, revealing that quasinormal modes are modified: damping rates generally decrease and ringing periods increase relative to GR, offering possible, but subtle, opportunities for gravitational-wave probes \cite{Churilova:2020bql,Ding:2018nhz}. Studies of quantum tunneling at both Killing and universal horizons in Einstein-Maxwell-æther black holes indicate nontrivial effects on Hawking radiation, especially under higher-order curvature corrections and modified dispersion relations \cite{Ding:2015fyx}.  The study of wormholes in Lorentz violating theories is also done in \cite{Ovgun:2018xys} where the authors obtained Exact traversable wormhole solution in bumblebee gravity.


The organization of this paper is as follows. In Sec. \ref{sec2}, we present a brief review of the EA theory and its corresponding field equations. In Sec. \ref{sec3}, we investigate traversable wormhole geometries within the framework of EA theory. By adopting an anisotropic form for the energy–momentum tensor, we reformulate the field equations accordingly. In Secs. \ref{sec4}, \ref{sec5}, and \ref{sec6}, we obtain solutions to these field equations for three different classes of combinations of the EA coupling constants. Within each class, we analyze three types of wormhole shape functions, and after determining the energy density and pressure components, we examine the energy conditions associated with the resulting wormhole solutions. Finally, in Sec. \ref{sec7}, we summarize and discuss our main findings.

\section{Einstein-Aether theory and Field equations}\label{sec2}

The action of EA theory is defined as \cite{Jacobson:2004ts,Ding:2015kba}  
\bea\label{action}
S=\int d^4x \sqrt{-g}\left[\frac{1}{16\pi G}\left(R+\mathcal{L}_{\ae}\right)+\mathcal{L}_M\right],
\eea
where
\bea
\mathcal{L}_{\ae}=-{K^{ab}}_{mn}\nabla_a u^m\nabla_b u^n+\l (g_{ab}u^a u^b+1),
\eea
and
\bea
 {K^{ab}}_{mn}=c_1g^{ab}g_{mn}+c_2\d^a_m\d^b_n+c_3\d^a_n\d^b_m-c_4u^au^bg_{mn}.
\eea
Here $c_i$'s are dimensionless aether coupling constants, $u^a$ is the aether field, $\l$ is a Lagrange multiplier enforcing the unit time-like constraint on the aether, and $\mathcal{L}_M$ is the matter Lagrangian which depends on the metric and matter fields. The Kronecker delta is defined as
\be
\d^a_m=g^{a \a}g_{\a m}\,.
\ee
Variation of the action (\ref{action}) with respect to $\l$ yields the constraint
\be \label{u1}
g_{ab}u^au^b=-1\,,
\ee
which ensures that the aether field is a unit time-like vector. In the weak-field and slow-motion limit, the EA theory reduces to Newtonian gravity, where the Newtonian constant $G_N$ is related to $G$ in (\ref{action}) as \cite{Garfinkle:2007bk}
\be
G=G_N\big(1-\frac{c_{14}}{2}\big)\,,
\ee
with $c_{ij}\equiv c_i+c_j$. Observational and theoretical considerations impose the following constraints on the coupling constants \cite{Jacobson:2007fh,Yagi:2013ava,Ding:2015kba}
\be\label{ccon}
0\le c_{14}<2\,, \qquad 2+c_{13}+3c_2>0\,, \qquad 0\le c_{13}<1\,.
\ee

Variation of the action (\ref{action}) with respect to the aether field $u^a$ gives the aether field equation \cite{Garfinkle:2007bk}
\be \label{eomu}
\nabla_aJ^a_b+c_4 a_a \nabla_b u^a +\l u_b=0,
\ee
where 
\be 
J^a_{~m}={K^{ab}}_{mn}\nabla_b u^n  , \qquad a_a=u^b\nabla_b u_a\,,
\ee
Variation of (\ref{action}) with respect to the metric $g_{\mu\nu}$ leads to \cite{Chan:2021ela,Ding:2015kba,Ding:2020bwa}
\be \label{eom}
\mathcal{G}_{ab}=T_{ab}^{\ae}+8\pi G T^M_{ab}\,,
\ee
where $\mathcal{G}_{ab}=R_{ab}-\frac{1}{2}R g_{ab}$ is the Einstein tensor, and the aether energy-momentum tensor is 
\bea  \label{eomg}
&&T_{ab}^{\ae}=\lambda u_au_b+c_4a_aa_b-\frac{1}{2}g_{ab}J^c_{~d}\nabla_cu^d+\nabla_cX^c_{~ab}
+c_1[(\nabla_au_c)(\nabla_bu^c)-(\nabla^cu_a)(\nabla_cu_b)],\nn\\
&& ~~~~~~~~~~~~ X^c_{~ab}=J^c_{~~(a}u_{b)}-u_{(a}J^{~~c}_{b)}+u^cJ_{(ab)}\, , \qquad \l=c_4 a^2+u^a\nabla_bJ^b_{~a}\,.
\eea
The matter energy-momentum tensor is also defined as
\be \label{eomg2}
T^M_{ab}=-\frac{2}{\sqrt{-g}}\frac{\delta (\sqrt{-g} \mathcal{L}_M)}{\delta g_{ab}}.
\ee

In \cite{Chan:2021ela} the field equations of EA theory are solved for  static and spherically symmetric spacetime
\be \label{ssmetr}
ds^2=-A(r)dt^2+B(r)dr^2+r^2d\theta^2+r^2\sin^2\theta d\phi^2,
\ee
In particular, the authors of \cite{Chan:2021ela}, obtained some classes of black hole solutions of the EA theory, for certain choices of the coupling constants $c_i$'s as I: $c_2\neq0$, $c_{13}\neq0$, and $c_{14}=0$, II: $c_2\neq0$, $c_{13}=0$, and $c_{14}=0$, III: $c_2=0$, $c_{13}\neq0$, and $c_{14}=0$, IV: $c_2=0$, $c_{13}=0$, and $c_{14}\neq0$, $ \cdots $.
In the next section, we focus on traversable wormhole solutions in EA theory and analyze their energy conditions.

\section{EA traversable wormhole geometries}\label{prop}\label{sec3}

Satisfying the energy conditions is one of the major challenges in wormhole physics. In many modified theories of gravity, obtaining wormhole solutions by assuming a specific equation of state and solving the corresponding field equations is highly nontrivial and may even be analytically intractable. An alternative approach, commonly adopted in the literature, is to consider prescribed wormhole geometries, such as the Morris-Thorne geometry, and investigate whether the corresponding energy conditions are satisfied in the framework of a given gravitational theory (see \cite{Singh:2020rai,Jusufi:2020yus,Samart:2021tvl} and \cite{Dehghani:2009zza}, and references therein). Following this approach, in the present work, we employ the wormhole geometries introduced in \cite{Dehghani:2009zza} to investigate the energy conditions for wormhole configurations in EA theory.

Teraversable wormhole solutions are introduced in \cite{Morris:1988cz}. In this article we consider the traversable wormholes with vanishing tidal force, in which the line element is in the form 
\be \label{wmetr}
ds^2=-dt^2+\frac{dr^2}{1-\frac{b(r)}{r}}+r^2\left(d\theta^2+\sin^2\theta d\phi^2\right)\,,
\ee
where $b(r)$ is the wormhole shape function, which satisfies the following conditions at the wormhole throat $r_0$ 
\bea \label{bc}
&&b(r_0)=r_0\,,\\
&&b'(r_0)<1\,.\label{bp}
\eea
To preserve the metric signature throughout the wormhole spacetime, the radial coordinate satisfies the bound $r>r_0$. It is also necessary that
\be \label{wsig}
1-\frac{b(r)}{r}>0\,.
\ee
To study wormhole solutions in the background of EA theory, we need to specify the aether field $u^a$. Due to the static and spherically symmetric structure of the metric (\ref{wmetr}), we adopt the following aether field ansatz \cite{Chan:2021ela}
\be \label{u2}
u^a=\left[h(r),a(r),0,0\right]\,.
\ee
Noting the constraint (\ref{u1}) it is straightforward to express $h(r)$ in terms of $a(r)$ as
\be \label{hr}
h(r)=\sqrt{\frac{r\,a(r)^2+g(r)}{g(r)}}\,,
\ee
where $g(r)=r-b(r)$.
At this stage, we consider the matter energy-momentum tensor in the diagonal form
	$ {T^a}_{b}{}^{M}=\mathrm{diag}[-\rho(r),p_r(r),p_t(r),p_t(r)] $,
	where $ \rho $ denotes the energy density, while $ p_r $ and $ p_t $ represent the radial and transverse pressures, respectively. Substituting the metric (\ref{wmetr}) and the aether field (\ref{u2}) into the field equations (\ref{eomu})–(\ref{eomg2}), we obtain expressions for $ \rho $, $ p_r $, and $ p_t $. We then reorganize these expressions in terms of the aether coupling constants $ c_i $ and their corresponding combinations, yielding:
\bea \label{eomro}
&& 	\begin{aligned}
		&\frac{c_2}{2\,r^2\,g^3(r)}
		\bigg\{	2r^3a^3 g^2(r)a''
		+r^3a^4 g(r)b''	-r^2 g^3(r)(a')^2\\
		&\qquad+r a g(r)\big[ra^2-g(r)\big]
		(rb'+4r-5b)a'
		+\frac{a^2}{4}\Big\{r^2\big[4ra^2-g(r)\big](b')^2
		\\
		&\qquad-2r\left[4a^2r^2+g(r)(4r-5b)\right]b'
		+4r(-4r^2+10br-5b^2)a^2
		-g(r)(4r-5b)^2\Big\}
		\bigg\}
	\end{aligned}\nn\\
&&	\begin{aligned}
	&+\frac{c_{13}}{2\,r^2\,g^3(r)}
		\bigg\{
		2r^3a^3g^2(r)a''
		+r^3a^4g(r)b''	-r^2g^3(r)(a')^2
		\\
		&\qquad+r a g(r) \Big\{r\big[ra^2-g(r)\big]b' +(4r^2-5rb)a^2	+b g(r)
		\Big\}a'
		+a^2\Big[
		r^2\Big(ra^2-\frac{g(r)}{4}\Big)(b')^2
		\\
		&\qquad-2br\Big(ra^2-\frac{g(r)}{4}\Big)b'	+r(-4r^2+8br-3b^2)a^2
		-2g(r)\Big(
		\frac{9b^2}{8}-2rb+r^2
		\Big)
		\Big]
		\bigg\}
	\end{aligned}\nn\\
&&	\begin{aligned}
		&-\frac{c_{14}}{2g^3(r)\big[ra^2+g(r)\big]r^2}
		\bigg\{	2ar^2g^2(r)
		\big[r^2a^4-g^2(r)\big]a''
		+ra^2g(r)
		\big[r^2a^4-g^2(r)\big]b''
		\\
		&\qquad+r^2g^3(r)
		\big[ra^2-2g(r)\big](a')^2+g(r)ra
		\Big\{r\big[r^2a^4+rg(r)a^2-3g^2(r)\big]b'
		\\
		&\qquad +(4r^3-5br^2)a^4-rbg(r)a^2 -g^2(r)(4r-7b)	\Big\}a'
		\\
		&\qquad+(rb'-b)^2a^2
		\bigg[
		r^2a^4+\frac{rg(r)a^2}{4}
		-\frac{3g^2(r)}{2}
		\bigg]
		\bigg\}-\frac{b'}{r^2}+\r=0\,,
	\end{aligned}
\eea
\bea \label{eompr}
&&	\begin{aligned}
		&\frac{-c_2}{2r^2g^3(r)}
		\bigg\{
		2ar^2g^2(r)\big[ra^2+g(r)\big]a''+r^2a^2g(r)\big[ra^2+g(r)\big]b''
		+r^2g^3(r)(a')^2
		\\
		&\qquad+r a g(r)\big[ra^2+2g(r)\big]	(rb'+4r-5b)a'
		+a^2\Big\{r^2\Big[ra^2+\frac{5g(r)}{4}\Big](b')^2
		\\
		&\qquad-2r\Big[a^2r^2+\frac{5bg(r)}{4}\Big]b'
		+r(-4r^2+10br-5b^2)a^2
		+\frac{5b^2g(r)}{4}
		\Big\}
		\bigg\}
	\end{aligned}\nn\\
&&\begin{aligned}
		&-\frac{c_{13}}{2r^2g^3(r)}	\bigg\{
		2ar^2g^2(r)\big[ra^2+g(r)\big]a''+r^2a^2g(r)\big[ra^2+g(r)\big]b''
		+r^2g^3(r)(a')^2
		\\
		&\qquad+ a r g(r)\Big\{
		r\big[ra^2+2g(r)\big]b'
		+(4r^2-5rb)a^2	+g(r)\left(4r-6b\right)
		\Big\}a'
		\\
		&\qquad+a^2\Big\{
		r^2\Big[ra^2+\frac54g(r)\Big](b')^2
		-2br\Big[ra^2+\frac54g(r)\Big]b'
		+r(-4r^2+8br-3b^2)a^2
		\\
		&\qquad-2g(r)\Big(\frac{3b^2}{8}-2rb+r^2\Big)
		\Big\}	\bigg\}
	\end{aligned}\nn\\
&&	\begin{aligned}
		&+\frac{c_{14}\,a^2}{r\big[ra^2\!+g(r)\big]g^3(r)}
		\bigg\{
		ar^2g^2(r)\big[ra^2\!+g(r)\big]a''
		+\frac{r^2a^2g(r)\big[ra^2\!+g(r)\big]}{2}b''
		\\
		&\qquad+\frac{3r^2g^3(r)}{2}(a')^2+\frac{rag(r)}{2}
		\Big\{
		r\big[ra^2+4g(r)\big]b'
		+(4r^2-5rb)a^2
		\\
		&\qquad	+4g(r)(r-2b)
		\Big\}a'+\frac{a^2}{2}(rb'-b)^2
		\Big[ra^2+\frac74g(r)\Big]
		\bigg\}-\frac{b}{r^3}-p_r=0\,,
	\end{aligned}
\eea
\bea \label{eompt}
&&	\begin{aligned}
		&c_2\,\frac{4r^2g^2(r)
			\left[2a a''\!+(a')^2\right]
			+4r^2a^2g(r)b''
			+8rag(r)(rb'\!+4r-5b)a' +5a^2(rb'\!-b)^2}{-8\,r^2g^2(r)}
	\end{aligned}\nn\\
&&+c_{13}\,\frac{4g^2(r)r^2(a')^2 +4rag(r)(rb'-4r+3b)a'
		+a^2(rb'-4r+3b)(rb'-b)}{8\,g^2(r)r^2}\nn\\
&&-c_{14}\,	\frac{a^2\left[2rg(r)a'+a(rb'-b)\right]^2}
		{8g^2(r)r\left[ra^2+g(r)\right]}+\frac{b-b'r}{2r^3}-p_t=0\,.
\eea
Here, a prime denotes differentiation with respect to $ r $ and $g(r)=r-b$. It is worth noting that, by setting all aether coupling constants $ c_i $ to zero in Eqs. (\ref{eomro})–(\ref{eompt}), one recovers the corresponding expressions for the energy density $ \rho $ and pressures $ p_r $ and $ p_t $ of wormhole geometries in Einstein gravity \cite{Morris:1988cz,Mehdizadeh:2024oyd}. Our goal is to investigate the energy conditions for wormhole solutions in this theory. To this end, we first obtain explicit expressions for the energy density $\rho$ and pressure components $p_r$ and $p_t$, and then examine the validity of the Weak Energy Condition (WEC) by simultaneously checking the following inequalities: EC1: $\rho>0$,~ EC2: $\rho+p_r>0$ and EC3: $\rho+p_t>0$. We also verify the NEC by examining EC2 and EC3. Note that if the WEC holds, the NEC is automatically satisfied.

The equations (\ref{eomro})-(\ref{eompt}) are solvable for certain values of the coupling constants $c_i$. Among these, we identify three classes of solutions in which the satisfaction of energy conditions imposes some limitations on the aether coupling constants. In other words, we focus on traversable wormholes in the EA theory that respect the energy conditions. Our calculations show that such wormholes exist if the coupling constants lie within specific intervals.  In the following sections, we study these classes of wormhole solutions by considering three types of shape functions $b(r)$ and investigate the corresponding energy conditions.

\section{Class I: $c_{13}=0$, $c_{14}=0$ and $c_2\neq 0$}\label{sec4}

By choosing the  coupling constants of the EA theory as $c_{13}=0, c_{14}=0$ and $c_2\neq 0$, the equations (\ref{eomro})-(\ref{eompt}) become solvable, yielding \cite{Chan:2021ela} \footnote{In \cite{Chan:2021ela} the equations of motion of the EA theory are solved for  static spherically symmetric black hole metric (\ref{ssmetr}) and the authors find metric functions $A(r)$ and $B(r)$ and the function $a(r)$ of the aether field. In this work, we are going to study the wormhole geometries (for which the metric is in spherically symmetric form) in the background of EA theory. So we encounter the same equations for the aether field as in \cite{Chan:2021ela}  .}
\be \label{ua}
a(r)=\frac{d}{r^2}.
\ee
Here, $d$ is an integration constant. The energy density and pressure components can now be obtained by substituting $c_{13}=0$, $c_{14}=0$ and (\ref{ua}) into the main equations (\ref{eomro})-(\ref{eompt}), yielding
\bea \label{ropa}
&&\begin{aligned}
		\rho=&\frac{1}{4g^3(r)r^9}
		\bigg\{-2\,c_2r^2g(r)b''\,d^4+\frac{c_2r^2}{2}
		\big[r^3g(r)-4d^2\big](b')^2 d^2+\Big\{
		4r^7g^3(r)
		\\
		&+c_2 r
		\Big[-r^3b g(r)-4bd^2+8rd^2\Big]d^2
		\Big\}b'
		\\
		&+c_2 \Big[
		\frac{r^3b}{2}g(r)(b-2r)
		+6b^2d^2-12rbd^2+8r^2d^2
		\Big]d^2
		\bigg\},
	\end{aligned}\nn\\
\nn\\
&&\begin{aligned}
		p_r=&{1\over 8g^3(r)r^9}
		\bigg\{
		-4c_2r^2g(r)\big[r^3g(r)+d^2\big]b''d^2
		-c_2r^2\big[5r^3g(r)+4d^2\big](b')^2d^2
		\\
		&+c_2r	\Big[6b^2r^3-22br^4-8d^2b+16r^5+16rd^2	\Big]b'd^2
		-8r^6bg^3(r)
		\\
		&+c_2r^3b(-11b^2+27rb-16r^2)d^2
		+4c_2b(3b-4r)d^4
		\bigg\},
	\end{aligned}\\
&&\begin{aligned}
		p_t=-\frac{1}{8g^2(r)r^6}
		\bigg\{
		4r^2c_2g(r)b''d^2
		+\Big[
		4r^3g^2(r)
		+c_2\left(5rb'+11b-16r\right)d^2
		\Big](rb'-b)
		\bigg\}.\nn
	\end{aligned}
\eea

It is worth noting that the temporal component of the aether vector, given by Eq. (\ref{hr}), diverges at the wormhole throat, $ r=r_0 $. A similar behavior is encountered in black-hole solutions of Einstein-aether (EA) theory \cite{Estrada:2025lgo,Chan:2021ela}, where the aether field diverges at the black hole horizon. In the wormhole case, this divergent behavior of the aether field at the throat also leads to divergences in the total energy-momentum tensor, $ T_{\mu\nu}^{\ae}+T_{\mu\nu}^{M} $, as can be seen by evaluating Eqs. (\ref{eomro})–(\ref{eompt}) at $ r=r_0 $. In the vicinity of the throat, where $ b(r)=r_0+\epsilon $, the energy density and pressure components given in Eqs. (\ref{ropa}) take the following form:
\bea
&&\r= \frac{b'}{r_0^2}+c_2\,\frac{d^2}{8r_0^6}+c_2 \mathcal{O}(\epsilon^{-3})\,,\nn\\
&& p_r=-\frac{b}{r_0^3}+c_2\,\frac{11d^2}{8r_0^6}+c_2\mathcal{O}(\epsilon^{-3})\,,\nn\\
&&p_t= \frac{b-b'r_0}{2r_0^3}+c_2\,\frac{11d^2}{8r_0^6}+c_2\mathcal{O}(\epsilon^{-2})\,.
\eea
It is worth to mention  that, apart from the divergent contributions, the aether energy-momentum tensor also provides finite contributions to the energy density and pressure at the wormhole throat, as represented by the terms proportional to \(c_2\) in the above expressions. Furthermore, the corresponding results in Einstein gravity can be recovered by setting \(c_2=0\) in these finite contributions.

In the following, we consider three types of the wormhole shape function $b(r)$: the power-law, logarithmic, and hyperbolic shape functions \cite{Dehghani:2009zza} and we investigate the energy conditions for wormholes in these cases.

\subsection{Wormholes with power law shape function}

As the first type, we choose the power-law shape function as \cite{Dehghani:2009zza}
\be \label{shape}
b(r)=r_0 \left(\frac{r_0}{r}\right)^n\,,
\ee
We aim to identify regions in the parameter space of the solutions where the energy conditions are satisfied. Noticing the observational and theoretical constraints (\ref{ccon}) together with the combinations of couplings in this class ($c_{13}=c_{14}=0$), one finds a bound on the couplings as $c_2\geq -\frac23$\,. Recall also that the shape function $b(r)$ must satisfy the conditions (\ref{bc})-(\ref{wsig}). It is straightforward to verify that, for the shape function (\ref{shape}) these conditions are satisfied if the bound $n>-1$ is respected. 
By substituting the shape function (\ref{shape}) into Eq. (\ref{ropa}), the corresponding expressions for EC1, EC2, and EC3 are obtained as
\bea \label{ec1power}
&&\begin{aligned}
		{\rm EC1}&={1\over
			8r^9\left(r_0^{1+n}-r^{1+n}\right)^3}
		\bigg\{
		\left(r_0^{1+n}-r^{1+n}\right)
		\Big[	c_2(1+n)^2r^3r_0^{2+2n}+24nr^7r_0^{2+2n}d^2
		\\
		&-8nr^{6-n}r_0^{1+n}
		\big( r_0^{2+2n}+r_0^{1+n}r^{1+n}	+r^{2+2n} \big)	\Big]
		\\
		&+4c_2d^4(1+n)r^nr_0^{1+n} \Big[(n+4)r^{1+n}-3r_0^{1+n}	\Big]
		\bigg\},
	\end{aligned}\nn\\
&&\begin{aligned}
	{\rm EC2}&=	\frac{1+n}{2r^9\left(r_0^{1+n}-r^{1+n}\right)^3}\bigg\{
		c_2 r_0^{1+n}
		\Big[3r_0^{1+n}-(n+4)r^{1+n}\Big]
		\Big[r^3\big(r_0^{1+n}-r^{1+n}\big)-2r^nd^2\Big]d^2
		\\
		&-2r^{6-n}r_0^{1+n}
		\big(r_0^{1+n}-r^{1+n}\big)^3
		\bigg\},
	\end{aligned}\nn\\
&&
	\begin{aligned}
		{\rm EC3}
		&=\frac{1}{2r^9
			\left(r_0^{1+n}-r^{1+n}\right)^3}
		\bigg\{
		c_2(1+n)r_0^{1+n}\left[3r_0^{1+n}-(n+4)r^{1+n}\right]
		\\
		&\times
		\left[
		r^3\left(r_0^{1+n}-r^{1+n}\right)
		-r^nd^2	\right]d^2 +(1-n)r^{6-n}r_0^{1+n}
		\left(r_0^{1+n}-r^{1+n}\right)^3
		\bigg\}.
	\end{aligned}
\eea
At this stage, considering the bounds $n>-1$ and $c_2\geq -\frac23$, we investigate the parameter space of the wormhole solutions of the EA theory, in order to find regions in which the EC1, EC2 and EC3 in (\ref{ec1power}) take positive values.

As mentioned above, the energy density and pressure components diverge at the wormhole throat. Consequently, the energy-condition expressions EC1, EC2, and EC3, defined in Eq. (\ref{ec1power}), exhibit the same divergent behavior. To demonstrate this explicitly, we perform a Taylor expansion of these expressions about the throat, $ r=r_0 $, obtaining:
\bea
&&{\rm EC1}=-\frac{c_2 d^4}{4r_0^7}(r-r_0)^{-3}+\mathcal{O}(r-r_0)^{-2},\nn\\
&&{\rm EC2}=-\frac{c_2 d^4}{2r_0^7}(r-r_0)^{-3}+\mathcal{O}(r-r_0)^{-2},\nn\\
&&{\rm EC3}=-\frac{c_2 d^4}{4r_0^7}(r-r_0)^{-3}+\mathcal{O}(r-r_0)^{-2}.
\eea
It is clear that the above expressions are positive for negative values of the aether coupling $ c_2 $. Therefore, the weak energy condition (WEC), and consequently the null energy condition (NEC), are satisfied in the vicinity of the wormhole throat for $ c_2<0 $.

\begin{figure}
	\begin{picture}(0,0)(0,0)
		\put(114,-192){(a)}
		\put(340,-193){(b)}
	\end{picture}
	\begin{center}
		\includegraphics[height=6.1cm,width=6.5cm]{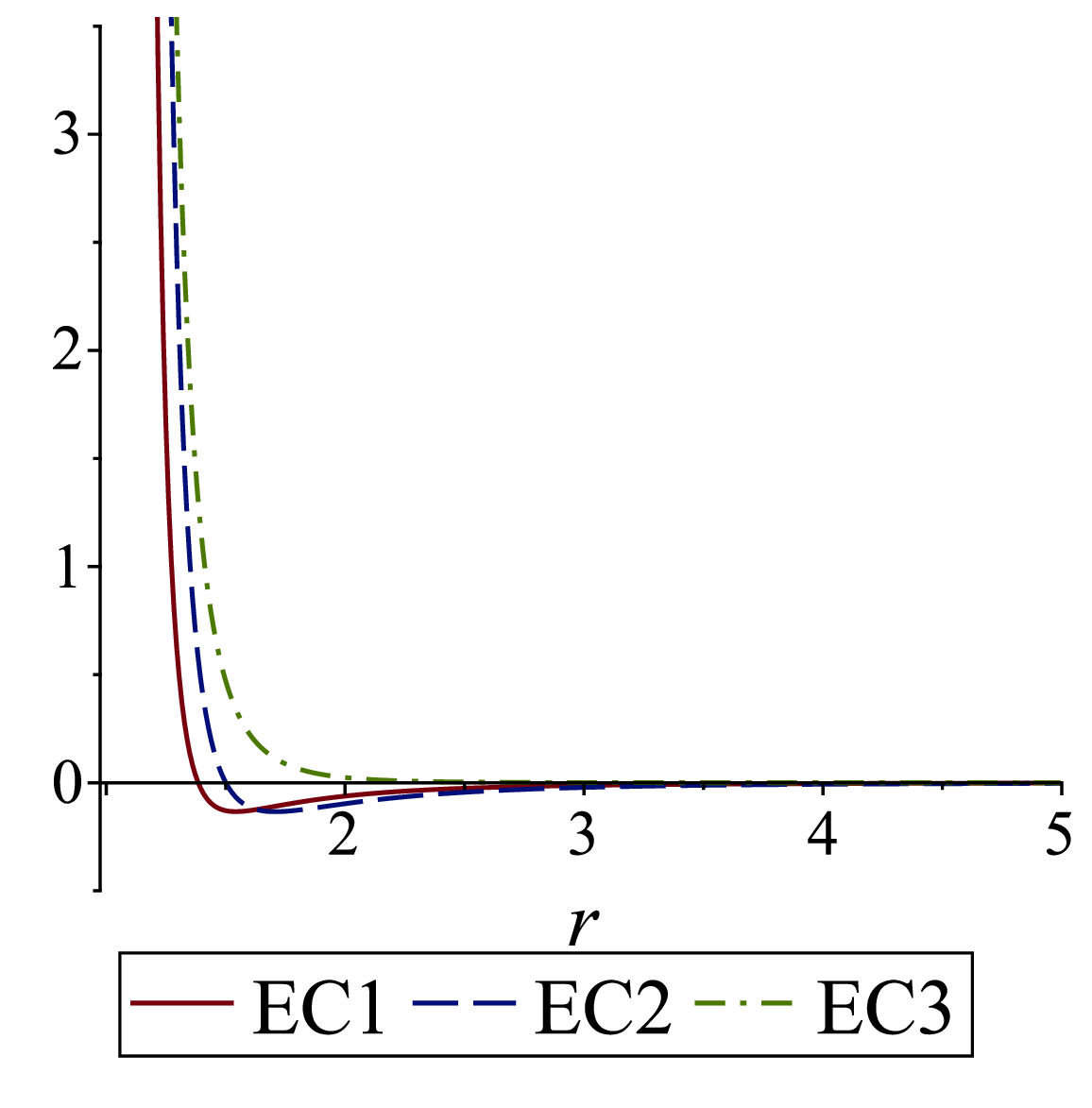}~~~~~~~ \includegraphics[height=6.1cm,width=6.5cm]{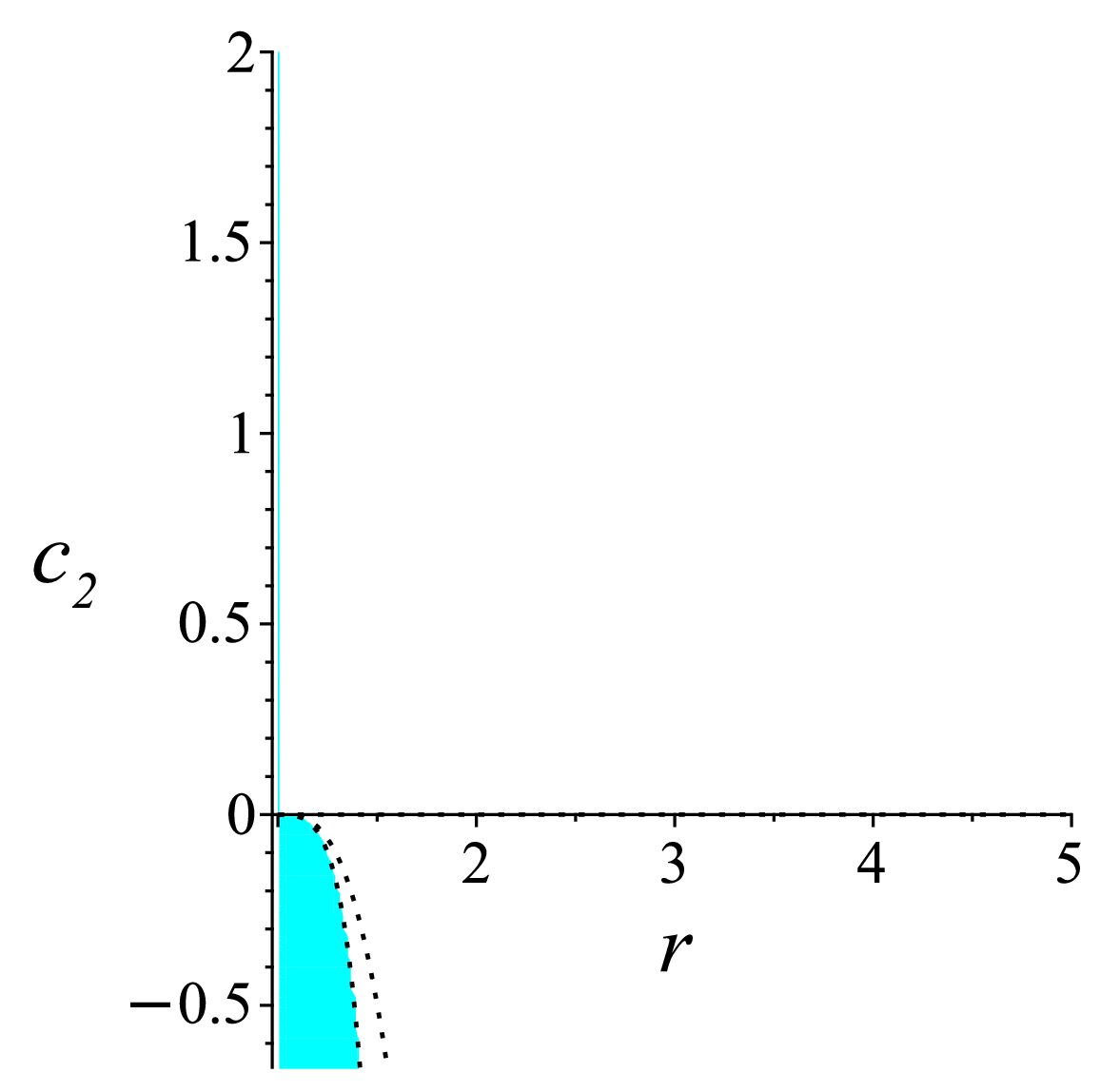}		
		\caption{\small Energy conditions for wormholes with power law shape function in the EA theory. Panel (a) is plotted by setting aether coupling $c_2=-\frac12$. We see that the energy conditions are respected around the wormhole throat (located at $r_0=1$). The blue region in panel (b) indicate where ${\rm EC1}>0$, ${\rm EC2}>0$, and ${\rm EC3}>0$ are satisfied simultaneously. This shows that WEC is respected in the throat's neighborhood and also short radial distances beyond that, if  $c_2$ takes adequate negative values. In these figures, we set $d=1$ and $n=1$.}\label{134rnc}
	\end{center}
\end{figure}

In Fig. \ref{134rnc}, we investigate the behavior of the energy conditions in the vicinity of the wormhole throat, located at $ r_0=1 $, and throughout the exterior region. In panel (a), we plot the energy-condition expressions for $ c_2=-\frac{1}{2} $. It is evident that EC1, EC2, and EC3 are all positive near the throat, indicating that both the WEC and NEC are satisfied in this region. In panel (b), we examine the dependence of the energy conditions on the aether coupling $ c_2 $ and the radial coordinate $ r $. The blue region represents the parameter space in which all three conditions, $ {\rm EC1}>0 $, $ {\rm EC2}>0 $, and $ {\rm EC3}>0 $, are simultaneously satisfied. As shown in this panel, for $ c_2<0 $, the WEC and NEC are satisfied over a finite radial interval extending beyond the wormhole throat.

\begin{figure}
	\begin{picture}(0,0)(0,0)
		\put(110,-200){(a)}
		\put(340,-200){(b)}
	\end{picture}
	\begin{center}
		\includegraphics[height=6.4cm,width=7cm]{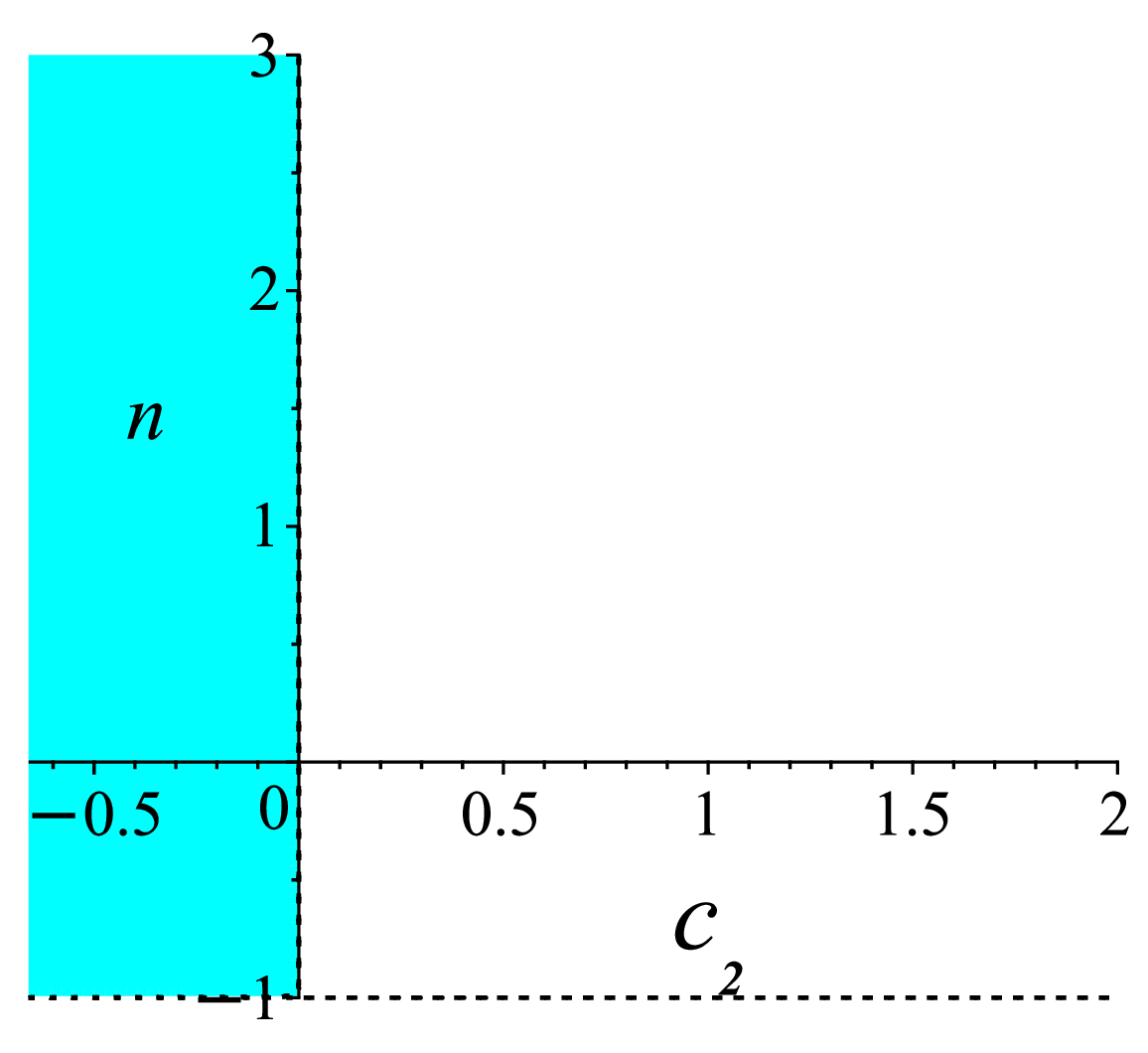}~~~ 	\includegraphics[height=6.4cm,width=7cm]{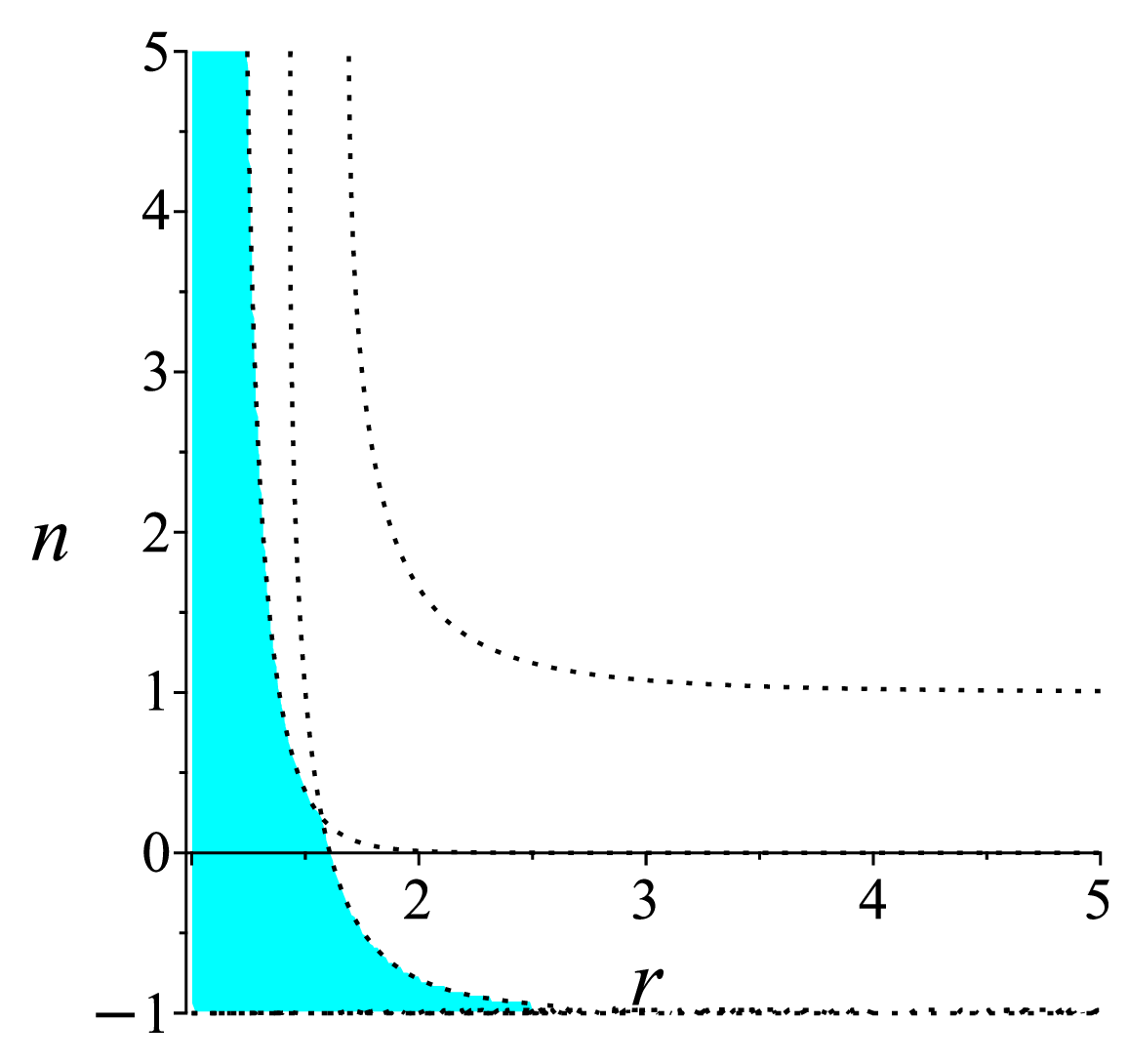}		
		\caption{\small Effect of exponent $n$ on the satisfaction of WEC for wormholes in EA theory. In panel (a) we set $r_0=1$ and fix the radial coordinate $r$ in the neighborhood of $r_0$. The blue color shows that WEC is satisfied in the throat's neighborhood if $c_2<0$, regardless of the values of $n$. Panel (b) which is plotted for $c_2=-\frac12$ shows that the width of satisfaction of WEC decreases by the growth  of exponent $n$. In these figures we also set $d=1$. }\label{134wecnec}
	\end{center}
\end{figure}

The influence of the exponent \(n\) on the energy conditions is investigated in Fig. \ref{134wecnec}. In panel (a), we consider the region in the vicinity of the wormhole throat, treating \(c_2\) and \(n\) as free parameters. The blue region, where the inequalities \({\rm EC1}>0\), \({\rm EC2}>0\), and \({\rm EC3}>0\) are simultaneously satisfied, indicates that the fulfillment of the WEC and NEC is largely insensitive to the value of \(n\), provided that the aether coupling \(c_2\) is negative. In panel (b), we fix \(c_2=-\frac{1}{2}\) and investigate the dependence of the energy conditions on \(n\) and \(r\). The blue region indicates where the WEC is satisfied. Our results show that, as \(n\) increases within the allowed range \(n>-1\), the radial interval over which the WEC is satisfied becomes narrower.

To construct the plots presented in this section and the following sections, our primary objective is to investigate how the model parameters affect the satisfaction of the energy conditions. For instance, the wormholes with  power law shape function, discussed in the above, are characterized by four parameters: \(n\), \(r_0\), \(d\) and \(c_2\). To obtain Fig. 2(a), we fix \(r_0=1\) and \(d=1\), while varying \(n\) and \(c_2\), in order to examine the satisfaction of the energy conditions. In some cases, choosing different values for the fixed parameters, such as \(r_0=10\) and \(d=5\), does not significantly affect the resulting plots. In other cases, however, changing the fixed parameter values may alter the size and shape of the blue region. Nevertheless, a consistent qualitative pattern emerges: for example, the blue region always remains in the vicinity of the wormhole throat. In other words, we have also investigated the behavior of the energy-condition expressions over parameter ranges and radial intervals beyond those displayed in the figures. Our results indicate that the overall qualitative behavior is largely insensitive to the specific values chosen for the fixed parameters.

\subsection{Wormholes with logarithmic shape function}
It is possible to consider the shape function in a logarithmic form \cite{Dehghani:2009zza}
\be \label{logb}
b(r)=r\, \frac{\ln r_0}{\ln r}\,.
\ee
Noticing the conditions (\ref{bc})-(\ref{wsig}), it is straightforward to show that the wormhole throat in this case must satisfy the bound $r_0>1$. By substituting the shape function (\ref{logb}) into  (\ref{ropa}), one can then obtain the corresponding expressions for the energy conditions as 
\bea \label{ecln}
&&\begin{aligned}
		\mathrm{EC1}
		=&\frac{c_2\ln r_0\, d^4}
		{2r^{10}(\ln r)^2(\ln r-\ln r_0)^3}	\left[-3(\ln r)^3
		+\big(3\ln r_0-2\big)(\ln r)^2	+\ln r_0\ln r\right]
		\\
		&+\frac{c_2r^4(\ln r_0)^2\,d^2}
		{8r^{10}(\ln r)^2(\ln r-\ln r_0)^2}+\frac{\ln r_0\,(\ln r-1)}{r^2(\ln r)^2}\,,
\end{aligned}\nn\\
&&\begin{aligned}
		\mathrm{EC2}=&-\frac{\ln r_0 c_2d^2}
		{r^6(\ln r)^2(\ln r-\ln r_0)^3}
		\Bigg[
		\frac{3}{2}(\ln r)^3
		+(1-3\ln r_0)(\ln r)^2+\frac{(\ln r_0)^2}{2}
		\\
		&	+\left(3(\ln r_0)^2-\frac{3}{2}\ln r_0\right)\ln r	\Bigg] 
		-\frac{\ln r_0}{r^2(\ln r)^2}
		\\
		&-\frac{\ln r_0c_2d^4}
		{r^{10}(\ln r)^2(\ln r-\ln r_0)^3}
		\left[
		3(\ln r)^3-3\ln r_0(\ln r)^2+(3(\ln r_0)^2-1)\ln r
		\right],
	\end{aligned}\nn\\
&&	\begin{aligned}
		\mathrm{EC3}
		=&\frac{\ln r_0}
		{2r^{10}(\ln r)^2(\ln r-\ln r_0)^3}
		\Big[
		r^8(\ln r-\ln r_0)^3(2\ln r-1)
		\\
		&+c_2d^2
		\left(
		r^4(\ln r-\ln r_0)
		+d^2\ln r
		\right)
		\left(
		3\ln r_0\ln r+\ln r_0
		-3(\ln r)^2-2\ln r
		\right)
		\Big].
	\end{aligned}
\eea

To examine the behavior of the energy-condition expressions in the vicinity of the wormhole throat, we expand them in a Taylor series about \(r=r_0\), obtaining:
\bea
&&{\rm EC1}=-\frac{c_2 \ln r_0\, d^4}{2r_0^7}(r-r_0)^{-3}+\mathcal{O}(r-r_0)^{-2},\nn\\
&&{\rm EC2}=-\frac{c_2 \ln r_0\, d^4}{r_0^7}(r-r_0)^{-3}+\mathcal{O}(r-r_0)^{-2},\nn\\
&&{\rm EC3}=-\frac{c_2 \ln r_0\, d^4}{2r_0^7}(r-r_0)^{-3}+\mathcal{O}(r-r_0)^{-2}.
\eea
Since \(r_0>1\), it is evident that the energy conditions are satisfied in the vicinity of the wormhole throat for \(c_2<0\).

\begin{figure}
	\begin{picture}(0,0)(0,0)
	\end{picture}
	\begin{center}
		\includegraphics[height=6.8cm,width=7.7cm]{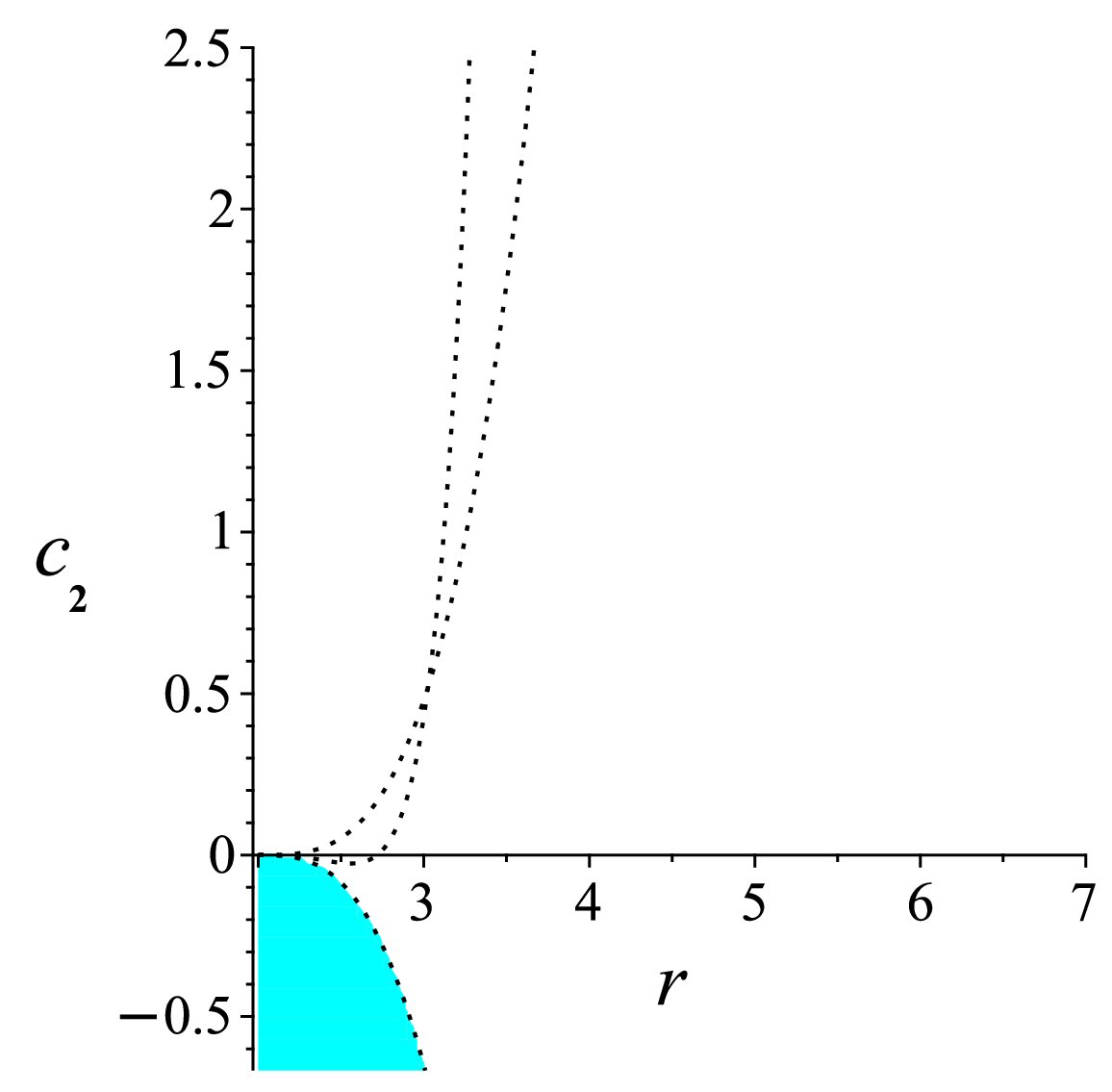}	
		\caption{\small Satisfaction of the energy conditions for the wormhole solutions with a logarithmic shape function in the EA theory. The blue region indicates that WEC (and also NEC) is satisfied around the wormhole throat, if $c_2<0$.  This figures is plotted for $d=3$ and $r_0=2$.}\label{lnwecnec}
	\end{center}
\end{figure}

Now to investigate the WEC and NEC, we examine whether there exist regions in the parameter space of these wormholes where the energy conditions (\ref{ecln}) are positive.
To further illustrate these results, we present Fig. \ref{lnwecnec}, where the blue region represents the simultaneous fulfillment of the inequalities \({\rm EC1}>0\), \({\rm EC2}>0\), and \({\rm EC3}>0\). As can be seen, both the WEC and NEC are satisfied in the vicinity of the wormhole throat for negative values of the aether coupling \(c_2\). We also investigate the region in parameter space where only \({\rm EC2}>0\) and \({\rm EC3}>0\) hold, corresponding to the satisfaction of the NEC without the WEC. Interestingly, we find that this region coincides with the blue region in Fig. \ref{lnwecnec}. 

\subsection{Wormholes with hyperbolic shape function}
\begin{figure}
	\begin{center}
		\includegraphics[height=7cm,width=8cm]{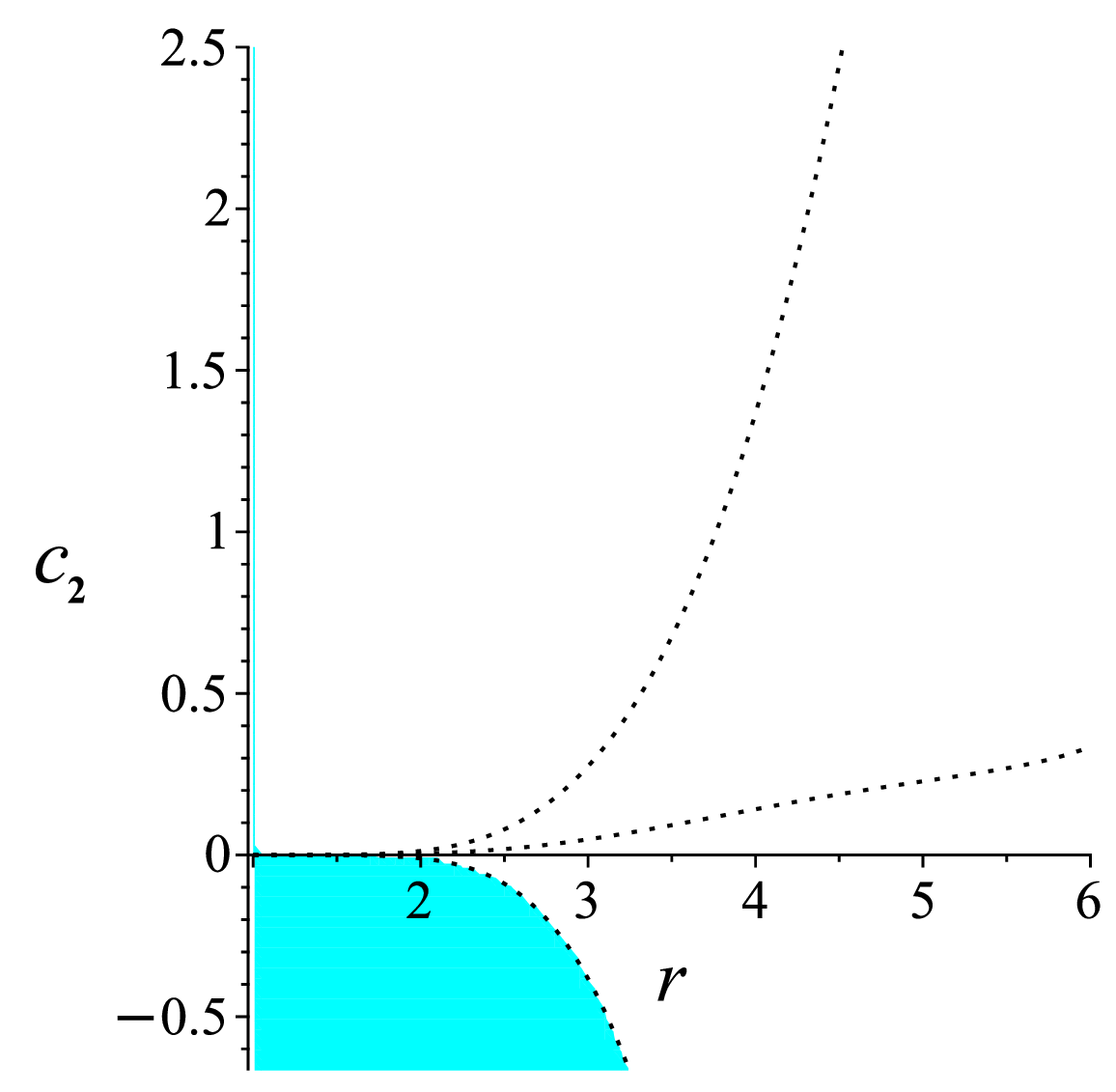}		
		\caption{\small Wormholes with the hyperbolic shape function in EA theory satisfy  WEC and NEC at the  vicinity of throat, if their parameters are set in the blue regions. In this figure we set $r_0=1$ and $d=5$.}\label{hpwecnec}
	\end{center}
\end{figure}

The last shape function considered in this section is the hyperbolic form. \cite{Dehghani:2009zza}
\be \label{hypb}
b(r)=r_0\, \frac{\tanh r}{\tanh r_0}\,.
\ee
In this case, all positive values of the wormhole throat radius $r_0>0$ satisfy the conditions (\ref{bc})-(\ref{wsig}). By substituting the above shape function into (\ref{ropa}), we obtain the following expressions for the energy conditions:
\bea
&&\!\!\!\!\begin{aligned}
		\mathrm{EC1}&=
		\frac{C_0 r_0}{r^9	\left(r_0 S C_0\!-r S_0 C\right)^3}
		\Bigg\{	r^3r_0^2	\bigg[	\frac{S^3c_2d^2}{8}	+\frac{S^2}{C}
		\Big(\frac{SC_0}{CS_0}\,r^3r_0
		-3r^4\!-\frac{c_2d^2}{4}\Big)r
		+\frac{Sc_2r^2d^2}{8C^2}
		\bigg]C_0^2
		\\
		&+3r_0S_0
		\Bigg[
		-\frac{d^2c_2 C(r^4+12d^2)S^2}{24}
		+r\left(r^8+\frac{r^4c_2d^2}{12}
		+\frac{c_2d^4}{3}\right)S
		\\
		&	-\frac{r^2c_2d^2
			\big(-8d^2C^2+(r^4+4d^2)\big)}{24C}
		\Bigg]C_0
		-r\left[
		d^4c_2(r^2-2C^2)S
		+rC(r^8+2c_2d^4)
		\right]S_0^2
		\Bigg\},
	\end{aligned}\nn\\
&&\!\!\!\!\begin{aligned}
		\mathrm{EC2}&=
		\frac{r_0C_0}
		{r^9S_0(r_0SC_0-rS_0C)^3}
		\Bigg\{\!
		-r\Bigg[\!
		\left(
		(-r^8\!-2r^4c_2d^2\!-4c_2d^4)C^2
		+c_2d^2r^2(r^4+2d^2)
		\right)S
		\\
		&		+r(r^8+2r^4c_2d^2+4c_2d^4)C
		\Bigg]S_0^3	+3\Bigg[	-\Big(r^8+\frac{7}{6}r^4c_2d^2+c_2d^4\Big)CS^2\\
		&	+r\Big(r^8+r^4c_2d^2+\frac{2}{3}c_2d^4\Big)S
		+\frac{2r^2c_2d^2}{3}
		\left(
		(r^4+d^2)C
		-\frac{1}{4C}
		\left(3r^4+2d^2\right)
		\right)
		\Bigg]C_0r_0S_0^2
		\\
		&-3C_0^2r^3r_0^2
		\Bigg[
		\left(-r^4-\frac{d^2c_2}{2}\right)S^3
		+r\left(
		\Big(r^4+\frac{c_2d^2}{3}\Big)\frac{S}{C}
		+\frac{rc_2d^2}{3}
		\right)S
		-\frac{Sc_2r^2d^2}{6C^2}
		\Bigg]S_0
		\\
		&+C_0^3r^6r_0^3\,\frac{S^3}{C}
		\left(\frac{r}{C}-S\right)
		\Bigg\},
	\end{aligned}\nn\\
&&\begin{aligned}
		\mathrm{EC2}&=
		\frac{r_0C_0}
		{2r^9S_0(r_0SC_0-rS_0C)^3}
		\Bigg\{\!	-r\Bigg[\!
		\left(	(r^8\!-4r^4c_2d^2-4c_2d^4)C^2
		+2c_2d^2r^2(r^4\!+d^2)	\right)S
		\\
		&+r(r^8+4r^4c_2d^2+4c_2d^4)C
		\Bigg]S_0^3
		+3C_0r_0
		\Bigg[
		C\left(r^8-\frac{7}{3}r^4c_2d^2
		-c_2d^4\right)S^2
		\\
		&	+r\Big(r^8\!+2r^4c_2d^2\!
		+\frac{2}{3}c_2d^4\Big)S
		+\frac{r^2c_2d^2}{3C}
		\Big(	\left(4r^4\!+2d^2\right)C^2
		-	\left(3r^4\!+d^2\right)
		\Big)
		\Bigg]S_0^2
		\\
		&-3C_0^2r^3r_0^2
		\Bigg[
		(r^4-d^2c_2)S^3
		+\left(
		\left(r^4+\frac{2d^2c_2}{3}\right)
		\frac{S}{C}	+\frac{2d^2rc_2}{3}
		\right)rS	-\frac{S d^2c_2r^2}{3C^2}
		\Bigg]S_0
		\\
		&+C_0^3r^6r_0^3\frac{S^3}{C}
		\left(\frac{r}{C}+S\right)
		\Bigg\}.
	\end{aligned}
\eea
Here, $S=\sinh r,\, C= \cosh r,\, S_0=\sinh r_0$ and $C_0= \cosh r_0 $. We performed a similar analysis and identified regions in the parameter space of these wormhole solutions where the energy conditions are satisfied. Our results, presented in Fig. \ref{hpwecnec}, show that the WEC, and consequently the NEC, can be satisfied in the vicinity of the wormhole throat for negative values of the aether coupling \(c_2\), as indicated by the blue region. A comparison of this figure with Figs. \ref{134rnc} and \ref{lnwecnec} reveals that wormholes with a hyperbolic shape function satisfy the WEC and NEC over a wider radial interval than those described by power-law and logarithmic shape functions. Nevertheless, the energy conditions are violated sufficiently far from the throat in all these wormhole configurations.
	
	We conclude this section by discussing the role of the aether coupling constant \(c_2\). Our results for wormhole geometries in Einstein-aether theory, with \(c_{13}=c_{14}=0\), indicate that the WEC and NEC are satisfied in the vicinity of the wormhole throat for negative values of \(c_2\). On the other hand, the constraints imposed by Eq. (\ref{ccon}) yield the lower bound \(c_2\geq -\frac{2}{3}\). Combining these results, we obtain the parameter range
	\be \label{conexo}
	-\frac{2}{3}\leq c_2<0,
	\ee
	which can be interpreted as a condition on the aether coupling for the existence of wormhole configurations satisfying the WEC and NEC near the throat, without requiring exotic matter in that region.

In addition to the constraints given in (\ref{ccon}), which lead to the condition (\ref{conexo}), gravitational-wave propagation in Einstein–Aether (EA) theory imposes a further restriction on the coupling constant \(c_2\), namely,
\[
\frac{c_{14}}{1-2c_{14}}<c_2,
\]
as discussed in \cite{Gong:2018cgj}. For the present class of solutions, where \(c_{14}=0\), this constraint reduces to
\begin{equation}
	c_2>0.
\end{equation}
This requirement rules out the possibility of satisfying condition (\ref{conexo}), which is necessary for avoiding exotic matter. Consequently, wormhole geometries in this class require exotic matter for their support, as is also the case in general relativity.

\section{Class II: $c_2=0$, and $c_{14}=0$}\label{sec5}

Another approach to solving the main equations (\ref{eomro})-(\ref{eompt}) is to set the coupling constants of the EA theory to $c_2=0$ and $c_{14}=0$. In this case, the aether field is expressed by the function \cite{Chan:2021ela}
\be \label{ub}
a(r)=\frac{e}{r^2}\,,
\ee
where $e$ is an integration constant.  By substituting $c_2=0$ and $c_{14}=0$ together with (\ref{ub}) into the field equations (\ref{eomro})-(\ref{eompt}), one obtains the expressions for the energy density $\rho$, and the radial and transverse pressure components $p_r$ and $p_t$, as
\bea \label{rpp214}
&&		\begin{aligned}
	\r=&\frac{1}{8r^9(r-b)^3}
	\bigg\{
	-4r^2e^4 c_{13}(r-b)b''
	+r^2e^2 c_{13}\left(r^4-r^3b-4e^2\right)
	\left(b'\right)^2
	\\
	&+8r^2b'\bigg[	r^5(r-b)^3
	+e^2 c_{13}\left(
	-\frac{3}{4}r^2b^2
	+\frac{7}{4}r^3b
	-r^4+e^2
	\right)
	\bigg]
	\\
	&+e^2 c_{13}\Big[
	-17r^3b^3
	+(57r^4+4e^2)b^2-(64r^5+8re^2)b	+24r^6
	\Big]
	\bigg\},
\end{aligned}\nn\\
&&	\begin{aligned}
	p_r
	=&\frac{e^2c_{13}}
	{8r^9(r-b)^3}
	\Bigg\{
	-4r^2(r-b)
	\left(r^4-br^3+e^2\right)b''
	-5r^2\Big(r^4-br^3+\frac{4e^2}{5}\Big)
	\left(b'\right)^2
	\\
	&
	+16r^2\Big(
	r^4\!-\frac{11br^3}{8}
	+\frac{3b^2r^2}{8}
	+\frac{e^2}{2}
	\Big)b'	-r^5(r-b)
	\left(13b^2-32br+24r^2\right)
	\\
	&
	+4e^2b\left(b-2r\right)
	\Bigg\}-\frac{b}{r^3},
\end{aligned}\\
&&\begin{aligned}
	p_t
	=&\frac{e^2c_{13}}
	{8r^6(r-b)^2}
	\Bigg[
	r^2\left(b'\right)^2
	-4r^2b'
	\left(-\frac{5}{2}b(r)+3r\right)
	+37b^2-84br+48r^2
	\Bigg]+\frac{b}{2r^3}\,.\nn
\end{aligned}
\eea
As discussed in the previous section, both the aether vector and the total energy-momentum tensor diverge at the wormhole throat, \(r=r_0=b(r_0)\). In the vicinity of the throat, we introduce the expansion \(r=r_0+\epsilon\), where \(0<\epsilon\ll1\). The energy density and pressure components then behave as follows:
\bea
&&\r=\frac{b'}{r_0^2}+c_{13}\frac{17e^2}{8r_0^6}+c_{13}\mathcal{O}(\epsilon^{-3})\,,\nn\\
&& p_r=-\frac{b}{r_0^3}-c_{13}\frac{13e^2}{8r_0^6}+c_{13} \mathcal{O}(\epsilon^{-3})\,,\nn\\
&&p_t=\frac{b-b'r_0}{2r_0^3}+c_{13}\frac{37e^2}{8r_0^6}+c_{13} \mathcal{O}(\epsilon^{-2})\,.
\eea
Although the divergence originates from the aether energy-momentum tensor, the latter also contributes finite terms to the energy density and pressure at the wormhole throat, namely, the terms proportional to \(c_{13}\). The corresponding results in Einstein gravity can be recovered by setting all the aether coupling constants \(c_i\) to zero in these finite contributions.

At this stage, we substitute three types of shape functions, power-law, logarithmic, and hyperbolic, into the energy density and pressures given in (\ref{rpp214}) and investigate the corresponding energy conditions for the wormhole solutions of the EA theory in this case. According to the constraints in Eq. (\ref{ccon}), the aether coupling combination \(c_{13}\) must lie within the range \(0<c_{13}<1\), a condition that is satisfied throughout our analysis.

\subsection{The power law shape function}

We begin our analysis in this section with the power-law shape function $b(r)=r_0(r_0/r)^n$. Remember also that  conditions (\ref{bc})-(\ref{wsig}) for the power law shape function lead to the bound $n>-1$. By inserting this function into  (\ref{rpp214}), the corresponding energy conditions can be obtained as follows:
\bea \label{pls}
&&	\begin{aligned}
	\!\!\!	{\rm EC1}&=\frac{1}{8r^6\left(r_0^{1+n}-r^{1+n}\right)^3}\Big\{-e^2 c_{13}(n^2-14n+57)r^{1+n}r_0^{2+2n}
	-4e^4c_{13}(1+n)r^{n-3}r_0^{2+2n}
	\\
	&-24nr^{5+n}r_0^{2+2n}+\left[e^2c_{13}(n^2-6n+17)+24nr^4\right]r_0^{3+3n}-8e^2c_{13}(n-8)r_0^{1+n}r^{2+2n}\\
	&+4e^4c_{13}(n+2)(n+1)r_0^{1+n}r^{2n-2}-24e^2c_{13}r^{3+3n}
	-8nr^{3-n}r_0^{4+4n}
	+8nr^{6+2n}r_0^{1+n}\Big\},
	\end{aligned}\nn\\
&&\begin{aligned}
	\!\!\!{\rm EC2}&=\frac{(1+n)}{2r^9\left(r_0^{1+n}-r^{1+n}\right)^3}
	\Big\{\left(e^2c_{13}(3+n)r^{n+4}
	+2c_{13}e^4r^n+6r^{n+8}\right)r_0^{2+2n}\\
	&+\left(-6r^7-c_{13}e^2r^3\right)r_0^{3+3n}-2r_0^{1+n}e^4c_{13}(2+n)r^{1+2n}-r_0^{1+n}e^2c_{13}(2+n)r^{5+2n}\\
	&+2r^{6-n}r_0^{4+4n}
	-2r^{9+2n}r_0^{1+n}\Big\},
\end{aligned}\nn\\
&&  \begin{aligned}
\!\!\!{\rm EC3}=&\frac{1}{4r^9\left(r_0^{1+n}-r^{1+n}\right)^3}\Big\{\big[
		-e^2c_{13}(n^2-18n+89)r^{n+4}
		+6(1-n)r^{n+8}\\
		&-2e^4c_{13}(1+n)r^n
		\big]r_0^{2+2n}+\left[e^2c_{13}(n^2-8n+27)+6r^4(n-1)\right]r^3r_0^{3+3n}\\
		&+2e^4c_{13}(n+2)(n+1)r_0^{1+n}r^{1+2n}-2(n-1)r^{6-n}r_0^{4+4n}\\
		&-2\left(5n-49\right)e^2c_{13}r_0^{1+n}r^{5+2n}+2(n-1)r_0^{1+n}r^{9+2n}
		-36e^2c_{13}r^{3n+6}\Big\}.
	\end{aligned}
\eea

To investigate the behavior of the above terms near the wormhole throat, we perform a Taylor expansion of Eq. (\ref{pls}) about \(r=r_0\), obtaining:
\bea
&&{\rm EC1}=-\frac{c_{13} e^4}{4r_0^7}(r-r_0)^{-3}+\mathcal{O}(r-r_0)^{-2},\nn\\
&&{\rm EC2}=-\frac{c_{13} e^4}{2r_0^7}(r-r_0)^{-3}+\mathcal{O}(r-r_0)^{-2},\nn\\
&&{\rm EC3}=-\frac{c_{13} e^4}{4r_0^7}(r-r_0)^{-3}+\mathcal{O}(r-r_0)^{-2}.
\eea
It is clear that the simultaneous positivity of EC1, EC2, and EC3 requires a negative value of the coupling combination \(c_{13}\). However, such values are excluded by the constraint (\ref{ccon}). Therefore, for wormhole geometries with a power-law shape function, neither the WEC nor the NEC is satisfied at the throat.
	
\begin{figure}
	\begin{picture}(0,0)(0,0)
				\put(118,-204){(a)}
				\put(342,-204){(b)}
			\end{picture}
	\begin{center}
	\includegraphics[height=6.5cm,width=7cm]{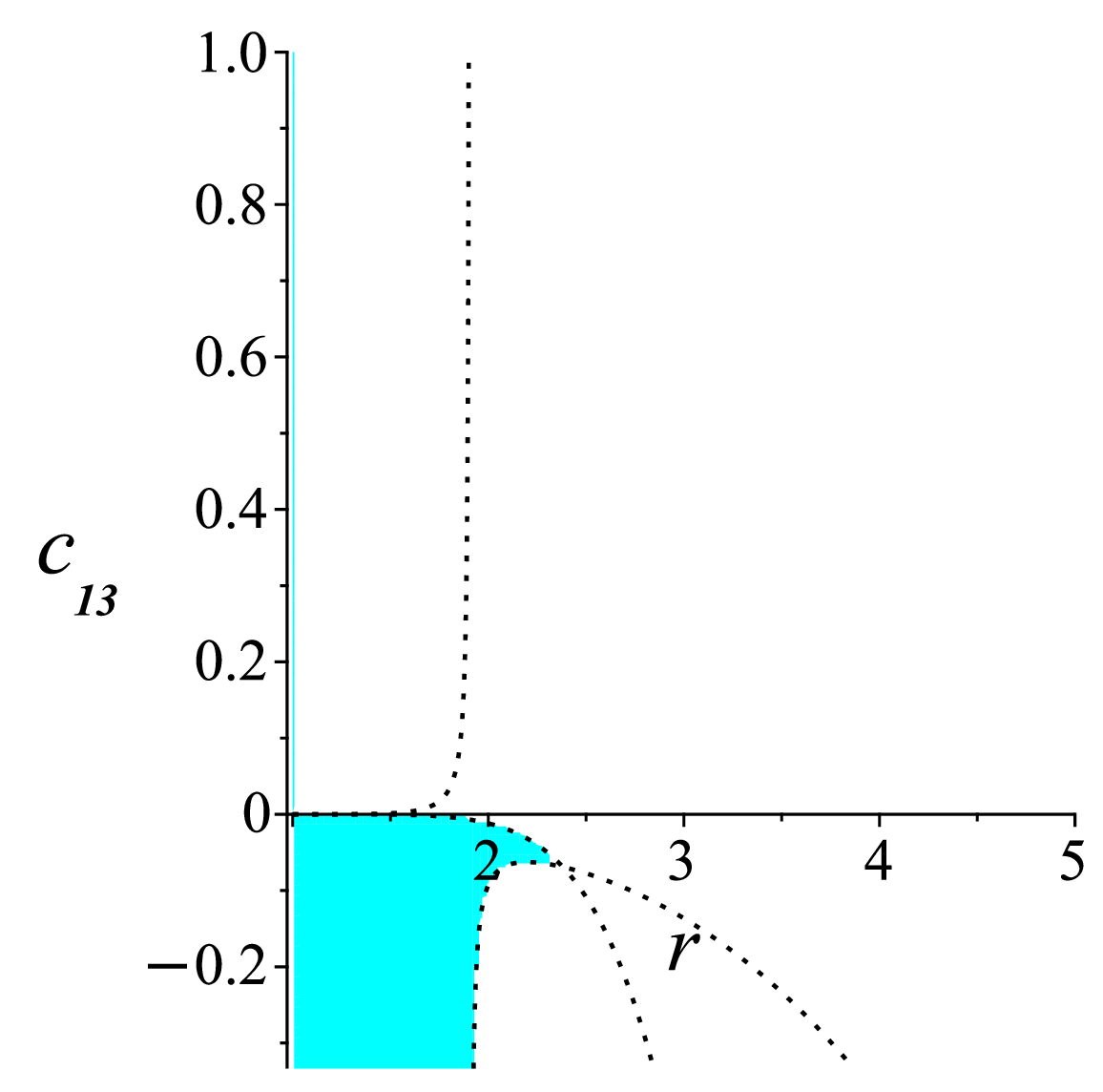}~~~~~
	\includegraphics[height=6.5cm,width=7cm]{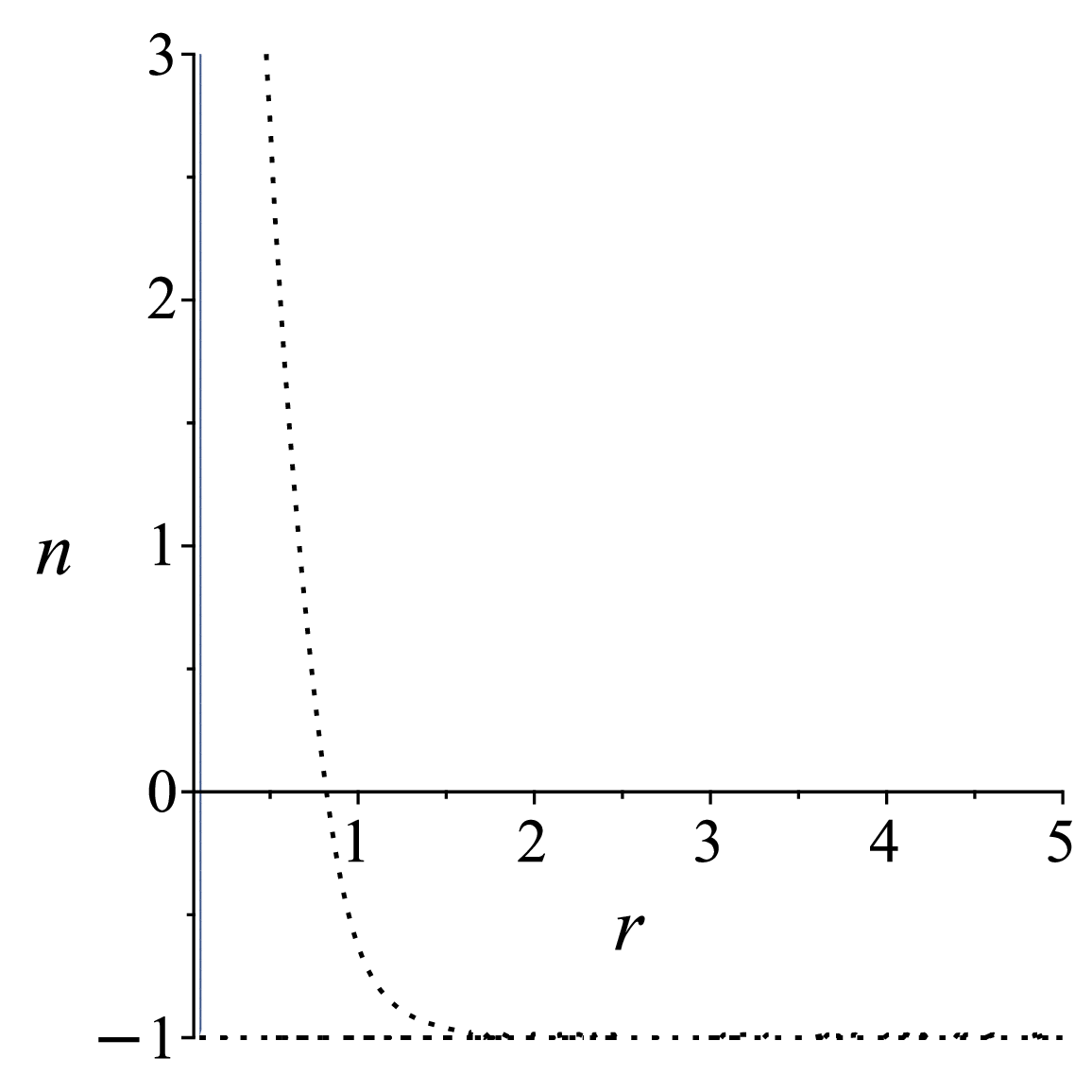}		
	\caption{\small The blue color indicates region in which ${\rm EC2}>0$ and ${\rm EC3}>0$ (means that NEC is respected) for the wormholes with power-law shape function. Panel (a) shows that the satisfaction of NEC may happens if $c_{13}<0$ which is not allowed by the constraint (\ref{ccon}). In panel (b) we set $c_{13}=0.5$ and vary the exponent $n$. The white color show that NEC is  nowhere respected. This figure is plotted by setting $r_0=1$ and $e=5$.}\label{214rnc}
	\end{center}
\end{figure}

	To investigate the energy conditions beyond the throat, we present Fig. \ref{214rnc}, which illustrates the dependence of EC2 and EC3 on the radial coordinate \(r\), for different values of the exponent \(n\) and the coupling \(c_{13}\). In panel (a), the blue region indicates that EC2 and EC3 can be simultaneously positive over a finite radial interval beyond the throat, located at \(r_0=1\), provided that \(c_{13}<0\). However, this range of \(c_{13}\) is excluded by the constraint (\ref{ccon}). Thus, the NEC, and consequently the WEC, cannot be satisfied throughout the spacetime for these wormhole configurations.
	
	In panel (b), we set \(c_{13}=\frac{1}{2}\), which lies within the allowed range, and examine the effect of the exponent \(n\) on the NEC beyond the throat. The absence of a blue region indicates that EC2 and EC3 are never simultaneously positive for the considered values of \(n\). Hence, neither the NEC nor the WEC is satisfied in the exterior region.


\subsection{The logarithmic shape function}

The energy conditions for the wormhole solutions in this case can be obtained by substituting the logarithmic shape function (\ref{logb}) into the energy density and pressure expressions (\ref{rpp214}). The result reads
\bea \label{eclog2}
&&\begin{aligned}
	\!\!\!\!\!{\rm EC1}&=\frac{1}{r^{10}(\ln r)^2\bigl(\ln r-\ln r_0\bigr)^3}\bigg\{
	-r^8\bigl(\ln r-1\bigr)(\ln r_0)^4+3e^2c_{13}r^4(\ln r)^5\\
	&+3r^4\left[
	(r^4-e^2c_{13})(\ln r)^2
	+\Big(-r^4+\frac{e^2c_{13}}{3}\Big)\ln r
	-\frac{e^2c_{13}}{24}
	\right](\ln r_0)^3\\
	&-3\!\left[
	(r^8\!-3e^2c_{13}r^4)(\ln r)^2\!
	+\Big(\frac{2}{3}e^2c_{13}r^4\!-\frac{1}{6}e^4c_{13}-r^8\Big)\ln r
	-\frac{e^2c_{13}(r^4\!+4e^2)}{24}
	\right]\ln r\,(\ln r_0)^2\\
	&+\left[
	(r^8-9e^2c_{13}r^4)(\ln r)^2
	+\Big(e^2c_{13}r^4-\frac{1}{2}e^4c_{13}-r^8\Big)\ln r
	-e^4c_{13}
	\right](\ln r)^2\ln r_0\bigg\},
\end{aligned}\nn\\
&&\begin{aligned}
	\!\!\!\!\!{\rm EC2}&=\frac{\ln r_0}{r^{10}(\ln r)^2\bigl(\ln r-\ln r_0\bigr)^3}\bigg\{
	\left(r^8+\frac{1}{2}e^2c_{13}r^4+e^4c_{13}\right)(\ln r)^3-r^8(\ln r_0)^3\\
	&+\left[
	\left(-3r^8-e^2c_{13}r^4-e^4c_{13}\right)\ln r_0
	+e^2c_{13}(r^4+2e^2)
	\right](\ln r)^2+\frac{1}{2}e^2c_{13}r^4(\ln r_0)^2\\
	&+3\ln r_0
	\left[
	\left(r^8+\frac{1}{6}e^2c_{13}r^4\right)\ln r_0
	-\frac{e^2c_{13}}{2}\left(r^4+\frac{2e^2}{3}\right)
	\right]\ln r\bigg\}, 
\end{aligned}\nn\\
&&\begin{aligned}
	\!\!\!\!\!{\rm EC3}&=\frac{\ln r_0}{2r^{10}(\ln r)^2\bigl(\ln r-\ln r_0\bigr)^3}\bigg\{-r^8\left(2\ln r-1\right)(\ln r_0)^4+18e^2c_{13}r^4(\ln r)^5\\
	&+6r^4\left[
	(r^4-3e^2c_{13})(\ln r)^2
	+\frac12\Big(\frac53e^2c_{13}-r^4\Big)\ln r
	-\frac{e^2c_{13}}{12}
	\right](\ln r_0)^3\\
	&-6\!\left[
	(r^8\!-9e^2c_{13}r^4)(\ln r)^2\!
	+\Big(\frac53e^2c_{13}r^4\!-\frac16e^4c_{13}-\frac{r^8}{2}\Big)\ln r
	-\frac{e^2c_{13}(r^4\!+2e^2)}{12}
	\right]\ln r\,(\ln r_0)^2\\
	&+\left[
	2(r^8-27e^2c_{13}r^4)(\ln r)^4
	+\left(5e^2c_{13}r^4-e^4c_{13}-r^8\right)(\ln r)^3
	-2e^4c_{13}(\ln r)^2
	\right]\ln r_0\bigg\}.
\end{aligned}
\eea
Here, the bound $r_0>1$ must be satisfied due to the conditions (\ref{bc})-(\ref{wsig}). We first perform a Taylor expansion of Eq. (\ref{eclog2}) about \(r=r_0\). As in the previous section, we find that the simultaneous positivity of EC1, EC2, and EC3 requires negative values of the coupling combination \(c_{13}\), which are excluded by the constraint (\ref{ccon}). Therefore, neither the WEC nor the NEC is satisfied at the throat of wormholes with a logarithmic shape function.
	
\begin{figure}
	\begin{center}
		\includegraphics[height=6.7cm,width=7.8cm]{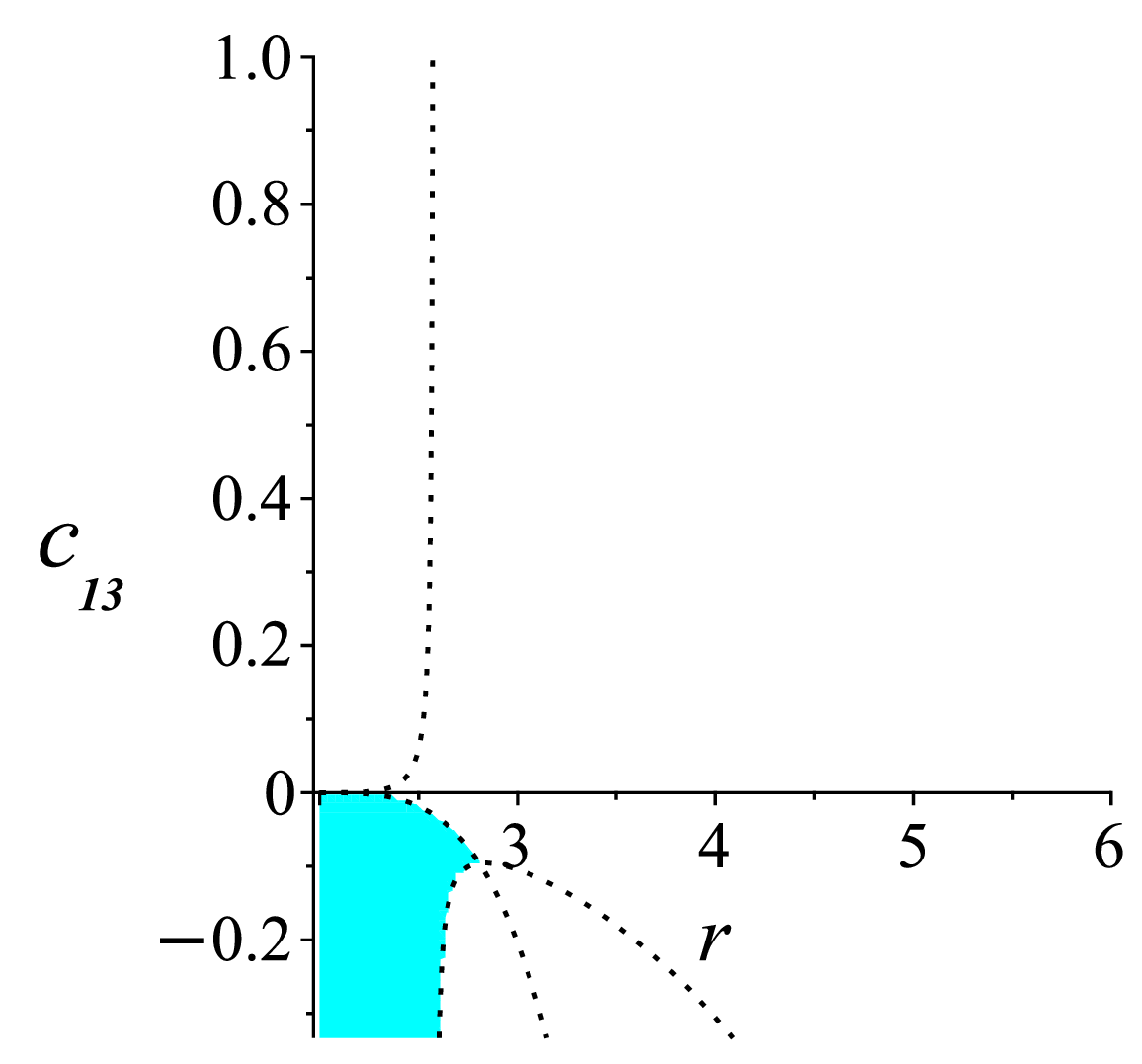}		
		\caption{\small Wormhole solutions of the EA theory with a logarithmic shape function does not satisfy the WEC and NEC, due to the constraint $0\leq c_{13}<1$. This figure is depicted for $r_0=2$ and $e=5$.} \label{214ln}
	\end{center}
\end{figure}

	We next investigate the energy conditions beyond the throat. To this end, we present Fig. \ref{214ln}, where the blue region represents the parameter space in which both \({\rm EC2}>0\) and \({\rm EC3}>0\) are satisfied. As can be seen, for positive (and hence allowed) values of \(c_{13}\), no such region exists. Thus, the NEC, and consequently the WEC, are violated throughout the exterior region. These results indicate that wormholes with a logarithmic shape function do not satisfy either the NEC or the WEC, either at the throat or beyond it.

\subsection{The hyperbolic shape function}

\begin{figure}
	\begin{center}
		\includegraphics[height=7cm,width=8.5cm]{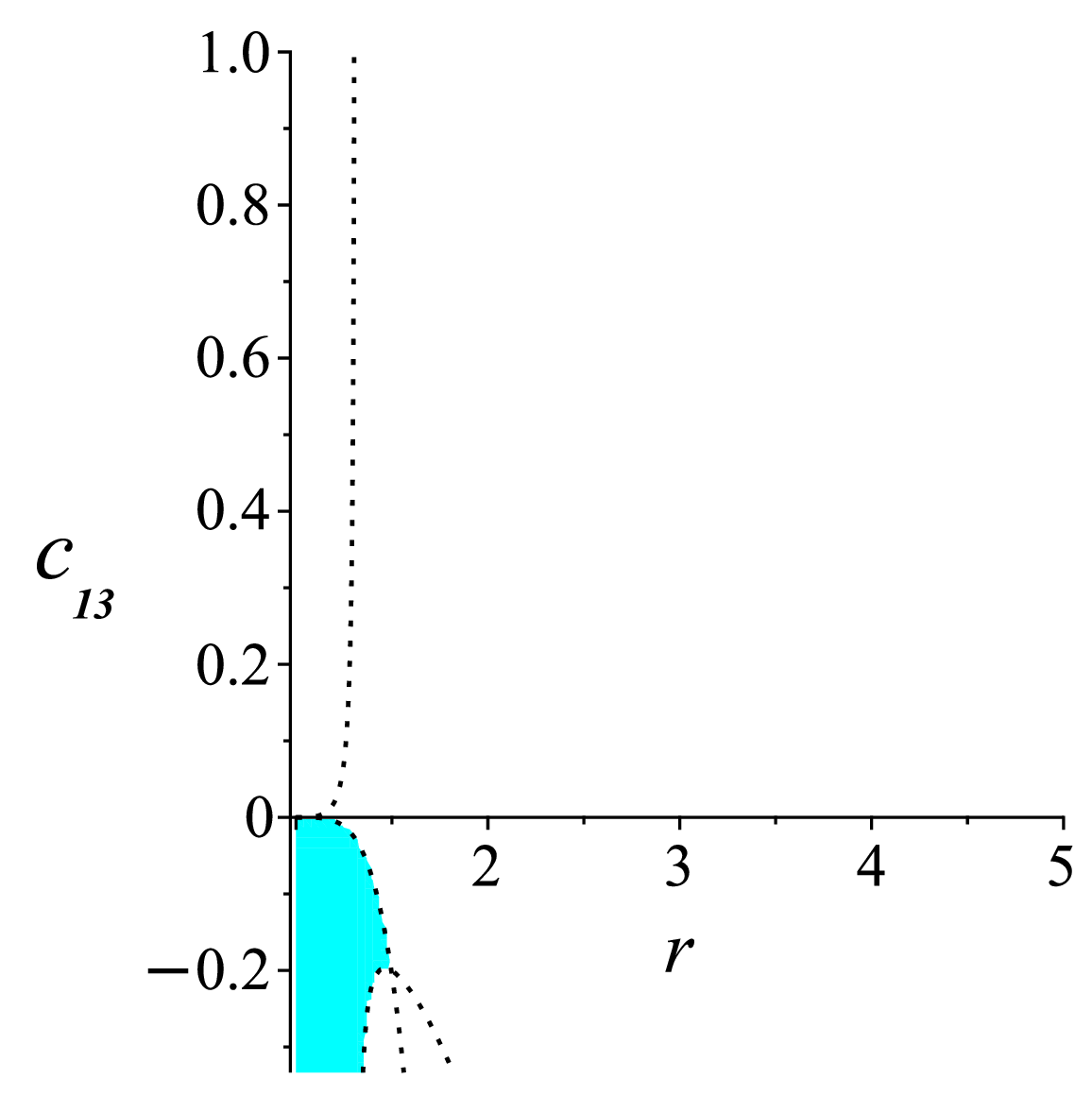}		
		\caption{\small The NEC and WEC are not respected by wormholes with hyperbolic shape function, since the blue region lies at $c_{13}<0$ which is rule out by the constraint (\ref{ccon}). In this figure we set $r_0=1$ and $ e=1$. } \label{hp214}
	\end{center}
\end{figure}

Considering the hyperbolic shape function for the wormhole solutions in this case, the energy conditions can be directly obtained by substituting (\ref{hypb}) into (\ref{rpp214}). The result is 
\bea
&&
	\begin{aligned}
\!\!\!\!{\rm EC1}&=\frac{1}{r^9\left(-r_0SC_0+rS_0C\right)^3}\Bigg\{3r^3r_0^3
	\bigg[
	-\frac{17e^2c_{13}S^3}{24}
	+r\frac{S}{C}
	\left(
	-\frac{r_0r^3S}{3}\frac{C_0}{S_0}
	+r^4-\frac{e^2c_{13}}{4}
	\right)\\
	&-\frac{r^2e^2c_{13}}{24}\frac{S}{C^2}
	\bigg]C_0^3-3\bigg[
	-\frac{19\left(r^4+\frac{4e^2}{57}\right)e^2c_{13}CS^2}{8}
	+\left(r^9-\frac{7}{12}r^5e^2c_{13}\right)S\\
	&	-\frac{e^2c_{13}r^2}{24}
	\left(-8e^2C+\frac{r^4+4e^2}{C}\right)
	\bigg]S_0r_0^2C_0^2+3S_0^3e^2c_{13}r^6C^3\\
	&+rS_0^2r_0
	\bigg\{
	-8e^2c_{13}
	\left[
	\left(r^4+\frac{e^2}{8}\right)C^2-\frac{r^2e^2}{8}
	\right]S
	+r(r^8-e^2c_{13}r^4+e^4c_{13})C
	\bigg\}C_0\Bigg\},
\end{aligned}\nn\\
&&\begin{aligned}
	\!\!\!\!{\rm EC2}&=\frac{r_0^2}{r^9\left(-r_0SC_0+rS_0C\right)^3}
	\Bigg\{
	rS_0^3\left[
	\left(-r^8-e^2c_{13}r^4-2e^4c_{13}\right)C^2
	+r^2e^2c_{13}(r^4+2e^2)
	\right]S\\
	&+r^2S_0^3
	\left(r^8+e^2c_{13}r^4+2e^4c_{13}\right)C -r^6r_0^3SC_0^3\frac{S}{C}
	\left(r\frac{S}{C}-S^2\right)   \\
	&-3r_0C_0S_0^2
	\Bigg[
	-\left(r^8+\frac12e^2c_{13}r^4+\frac13e^4c_{13}\right)CS^2
	+\left(r^9+\frac13r^5e^2c_{13}\right)S\\
	&+\frac{2e^2c_{13}r^2}{3}
	\left((r^4+e^2)C-\frac{3}{4C}\left(r^4+\frac{2e^2}{3}\right)\right)
	\Bigg]\\
	&+3r^3r_0^2C_0^2S_0
	\Bigg[
	-\left(r^4+\frac{e^2c_{13}}{6}\right)S^3
	+r^2\left(\frac{r^3S}{C}+\frac{e^2c_{13}}{3}\right)S
	-\frac{r^2e^2c_{13}S}{6C^2}
	\Bigg]
	\Bigg\}\frac{C_0}{S_0},
\end{aligned}\nn\\
&&\begin{aligned}
	\!\!\!\!{\rm EC3}&=\frac{1}{4r^9\left(-r_0SC_0+rS_0C\right)^3}
	\Bigg\{6r^3r_0^3C_0^3
	\Bigg[
	\left(
	-\frac{r_0r^3}{3}\frac{S}{C}\frac{C_0}{S_0}
	+r^4-\frac{9}{2}e^2c_{13}
	\right)S^3\\
	&+
	r\frac{S}{C}
	\left(
	-\frac{r_0r^3}{3}\frac{S}{C}\frac{C_0}{S_0}
	+r^4-\frac{4}{3}e^2c_{13}
	\right)S
	-\frac{r^2e^2c_{13}S}{6C^2}
	\Bigg]\\
	&-6r_0^2S_0C_0^2
	\Bigg[
	C\left(r^8-\frac{89}{6}e^2c_{13}r^4-\frac13e^4c_{13}\right)S^2
	+\left(r^9-3r^5e^2c_{13}\right)S\\
	&-\frac{e^2c_{13}r^2}{6}
	\left(-4e^2C+\frac{r^4+2e^2}{C}\right)
	\Bigg]+2r^2r_0S_0^2C_0
	\bigg[\Big[
	(r^8-49e^2c_{13}r^4-2e^4c_{13})C^2\\
	&	+2c_{13}e^4r^2
	\Big]S+r(r^8-5e^2c_{13}r^4+2e^4c_{13})C
	\bigg]+36e^2c_{13}r^6S_0^3C^3\Bigg\},
\end{aligned}
\eea
where $S=\sinh r, C=\cosh r, S_0=\sinh r_0$ and $C_0=\cosh r_0$.
Remember that in this case $r_0>0$ must be satisfied due to the conditions (\ref{bc})-(\ref{wsig}). To examine the validity of the NEC, we identify the region in which both \({\rm EC2}>0\) and \({\rm EC3}>0\) are satisfied. The corresponding results are presented in Fig. \ref{hp214}. It is evident that, as in the cases of power-law and logarithmic shape functions, wormholes with a hyperbolic shape function do not satisfy either the NEC or the WEC within the allowed range of the coupling parameter, \(0\leq c_{13}<1\).

We conclude this section by briefly summarizing our results. We have derived the energy density and pressure components for wormhole geometries in Einstein-aether (EA) theory in the case \(c_2=c_{14}=0\). Our analysis shows that wormholes with power-law, logarithmic, and hyperbolic shape functions can satisfy the NEC and WEC only for negative values of \(c_{13}\). However, such values are excluded in EA theory by the constraint (\ref{ccon}), which is imposed based on theoretical and observational considerations. We therefore conclude that, for \(c_2=c_{14}=0\), wormholes in EA theory require *exotic matter* for their support, as in Einstein gravity.

\section{Class III: $c_2=-c_{13}\neq0$, and $c_{14}=0$}\label{sec6}

It is also possible to solve the main equations (\ref{eomro})-(\ref{eompt}) by setting the coupling constants as $c_2=-c_{13}\neq0$ and $c_{14}=0$. with this choice the aether field can be obtained as \cite{Chan:2021ela}
\be
a(r)=\frac{1}{c_{13} r}\sqrt{c_{13} r(jc_{13}+r)}\,,
\ee
where $j$ is an integration constant. By substituting the above function and the values $c_2=-c_{13}$ and $c_{14}=0$ into the main equations, we obtain the corresponding energy density and pressure components as
\bea \label{rpp3}
&&\begin{aligned}
\r&=\frac{1}{	c_{13}r^3(r-b)^2}\Big\{r\left[
-c_{13}b^2
+\left(c_{13}r-jc_{13}^2\right)b
+r^2+jc_{13}(c_{13}+2)r
+c_{13}^2j^2
\right]b'\\
&+\left(jc_{13}^2+2c_{13}r\right)b^2+\left[
-(3c_{13}+1)r^2
-jc_{13}(c_{13}+2)r
-c_{13}^2j^2
\right]b
+c_{13}r^3\Big\},
\end{aligned}\nn\\
[3mm]
&&p_r=-\frac{
	r(jc_{13}+r)^2b'
	+c_{13}b^3
	-3c_{13}rb^2
	+\left[(3c_{13}-1)r^2-2jc_{13}r-c_{13}^2j^2\right]b
	-c_{13}r^3}{c_{13}r^3(r-b)^2},\nn\\
&&\begin{aligned}
  p_t&=\frac{1}{4r^3(r-b)^2}\bigg\{2r^2(r-b)(jc_{13}+r)b''\\
  &+3(rb'-b)\Big[
  r(jc_{13}+r)b'
  -\frac{2}{3}b^2
  +\Big(r+\frac{2jc_{13}}{3}\Big)b
  -\frac{4r}{3}\Big(r+\frac{5jc_{13}}{4}\Big)
  \Big]\bigg\}.
\end{aligned}
\eea
Similar to the previous sections, we now consider three different shape functions for the wormholes of the EA theory in this class and examine the energy conditions for them. Recalling the constraint (\ref{ccon}) among the coupling constants,  it is straightforward to see that the bound $-1<c_2$ must be satisfied for the solutions in this class and this restriction has been taken into account in the analysis and results that presented in this section.

\subsection{The power law shape function}
\begin{figure}
\begin{picture}(0,0)(0,0)
	\put(105,-209){(a)}
	\put(328,-209){(b)}
\end{picture}
	\begin{center}
		\includegraphics[height=6.7cm,width=7.4cm]{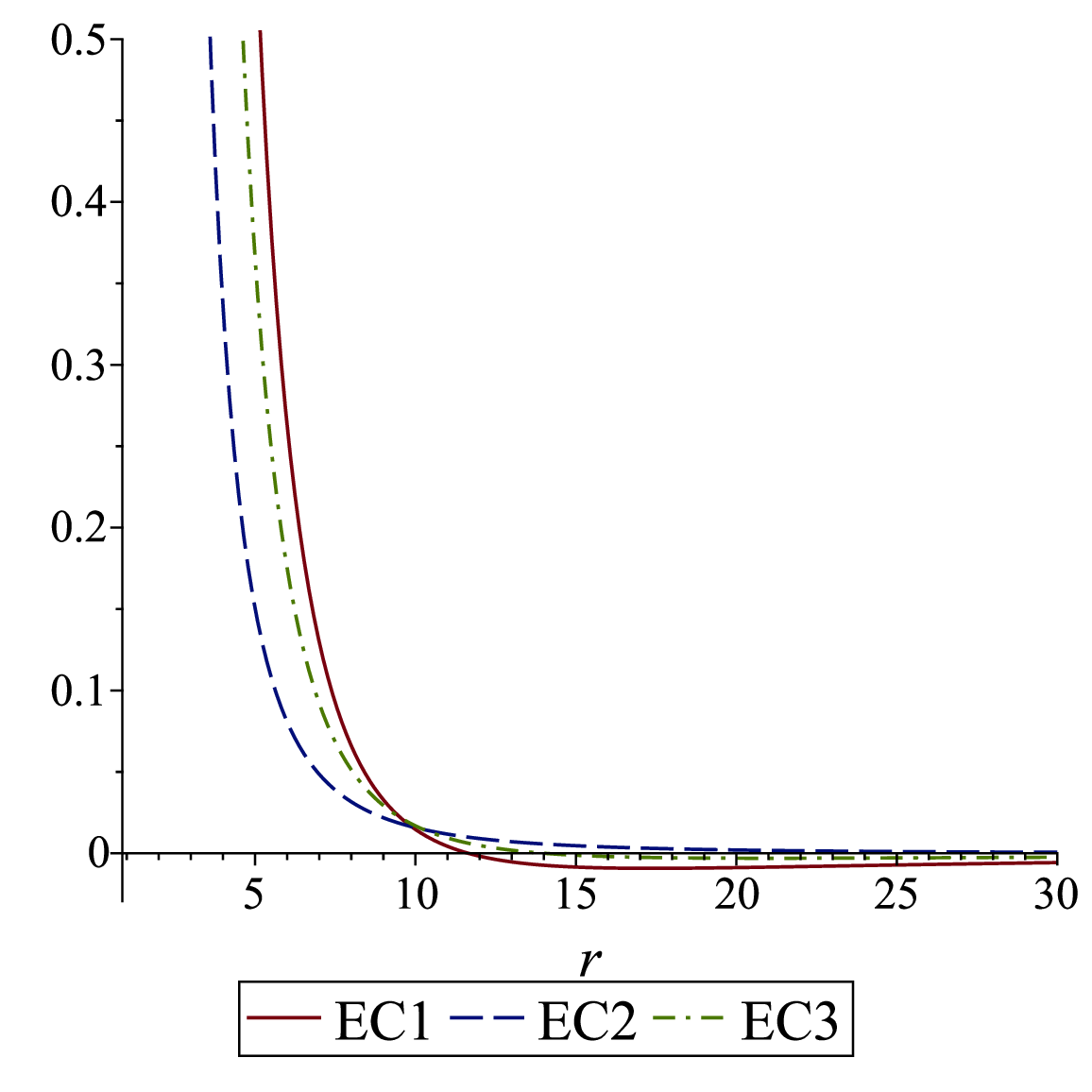}~~~~~
		\includegraphics[height=6.7cm,width=7.8cm]{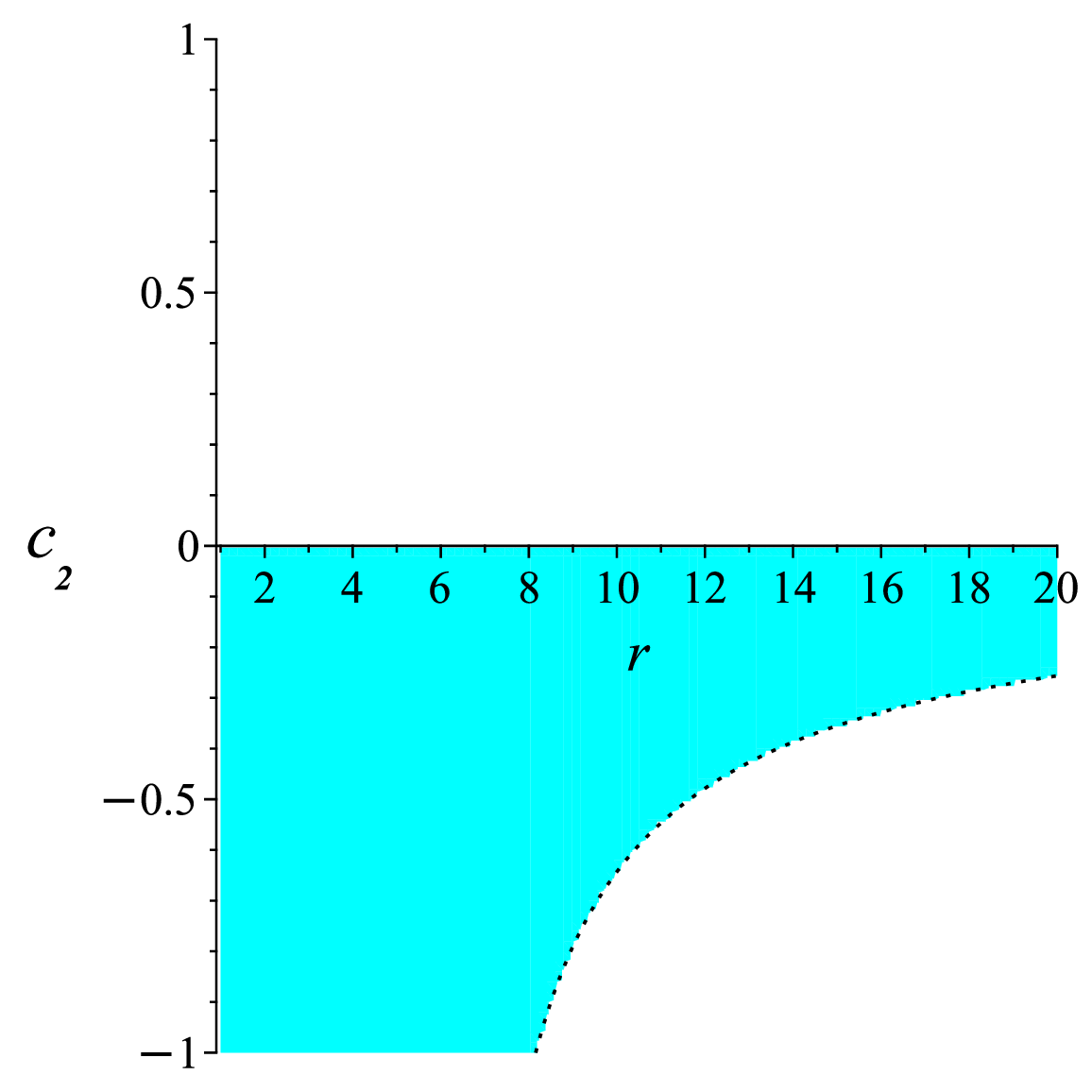}		
		\caption{\small Panel (a) which is depicted for $c_{2}=-\frac12$, indicates that energy condition take positive values near the wormhole throat ($r_0=1$). Blue color in panel (b) shows the domain where WEC is satisfied.	One can see that the energy conditions are respected throughout the entire spacetime for the wormholes with a power-law shape function in this class, provided that the combination $ c_{2} $ takes appropriate negative values. In this figure, we set $r_0=1, n=-1/2$ and $j=1$.} \label{rcnecwec}
	\end{center}
\end{figure}

In the case of wormhole solutions with the power-law shape function (\ref{shape}), and taking into account the energy density and pressure components given in (\ref{rpp3}), one can express the energy conditions as follows
\bea \label{ecp3}
&&\begin{aligned}
	\!\! {\rm EC1}&=\frac{1}{c_{2}r^3\left(r_0^{1+n}-r^{1+n}\right)^2}\bigg\{c_{2} \left[j(1+n)c_{2}+(n-2)r\right]r_0^{2+2n}-nc_{2}r^{-n}r_0^{3+3n}
	\\
	&-c_{2}r^{3+2n}-\Big[
	jc_{2}(c_{2}-2)(1+n)r^{1+n}
	+(n-3c_{2}+1)r^{2+n}
	+j^2c_{2}^2(1+n)r^n
	\Big]r_0\bigg\},
\end{aligned}\nn\\
&&\begin{aligned}
\!\! {\rm EC2}&=\frac{1+n}{c_2r^3\left(r_0^{1+n}-r^{1+n}\right)^2}
\Big\{\left[jc_2(4-c_2)r^{1+n}
-2j^2c_2^2r^n
-2r^{2+n}
\right]r_0^{1+n}
\\
&+\left(jc_2^2+c_2r\right)r_0^{2+2n}
-c_2r^{-n}r_0^{3+3n}
\Big\}.
\end{aligned}\nn\\
&&\begin{aligned}
	\!\! {\rm EC3}&= \frac{1}{4c_{2}r^3
		\left(r_0^{1+n}-r^{1+n}\right)^2}
	\Big\{-4j^2c_{2}^2(1+n)r^n r_0^{1+n}
	+jc_{2}(1+n)
	\Big[
	(6-n)c_{2}r_0^{2+2n}
	\\
	&	+\left(8-\left(2n+9\right)c_{2}\right)
	r^{1+n}r_0^{1+n}
	\Big]
	+c_{2}r(n^2+2n-11)r_0^{2+2n}
	\\
	&-2c_{2}(n-1)r^{-n}r_0^{3+3n}-4c_{2}r^{3+2n}
	+2\left[(n^2+3n+8)c_{2}
	-2(1+n)\right]r^{2+n}r_0^{1+n}
	\Big\}.
\end{aligned} 
\eea
Although the energy-condition expressions diverge at the wormhole throat, their behavior in its vicinity and throughout the exterior region can still be investigated. Our analysis shows that EC1, EC2, and EC3 can all take positive values near the throat and beyond it for negative values of the coupling \(c_{2}\). We illustrate the behavior of the energy conditions given in Eq. (\ref{ecp3}) in Fig. \ref{rcnecwec}. In panel (a), all three expressions are positive in the vicinity of the throat and exhibit divergent behavior as the throat is approached. The blue region in panel (b) represents the parameter space in which \({\rm EC1}>0\), \({\rm EC2}>0\), and \({\rm EC3}>0\) are simultaneously satisfied. This shows that the WEC, and consequently the NEC, can be satisfied within the allowed range \(-1< c_{2}<0\). An important feature of this class of wormhole solutions is that, for sufficiently small negative values of \(c_{2}\), the energy conditions can be satisfied not only in the vicinity of the throat, located at \(r_0=1\), but throughout the entire spacetime.

\begin{figure}
	\begin{picture}(0,0)(0,0)
		\put(105,-198){(a)}
		\put(340,-198){(b)}
		\end{picture}
	\begin{center}
		\includegraphics[height=6.3cm,width=7cm]{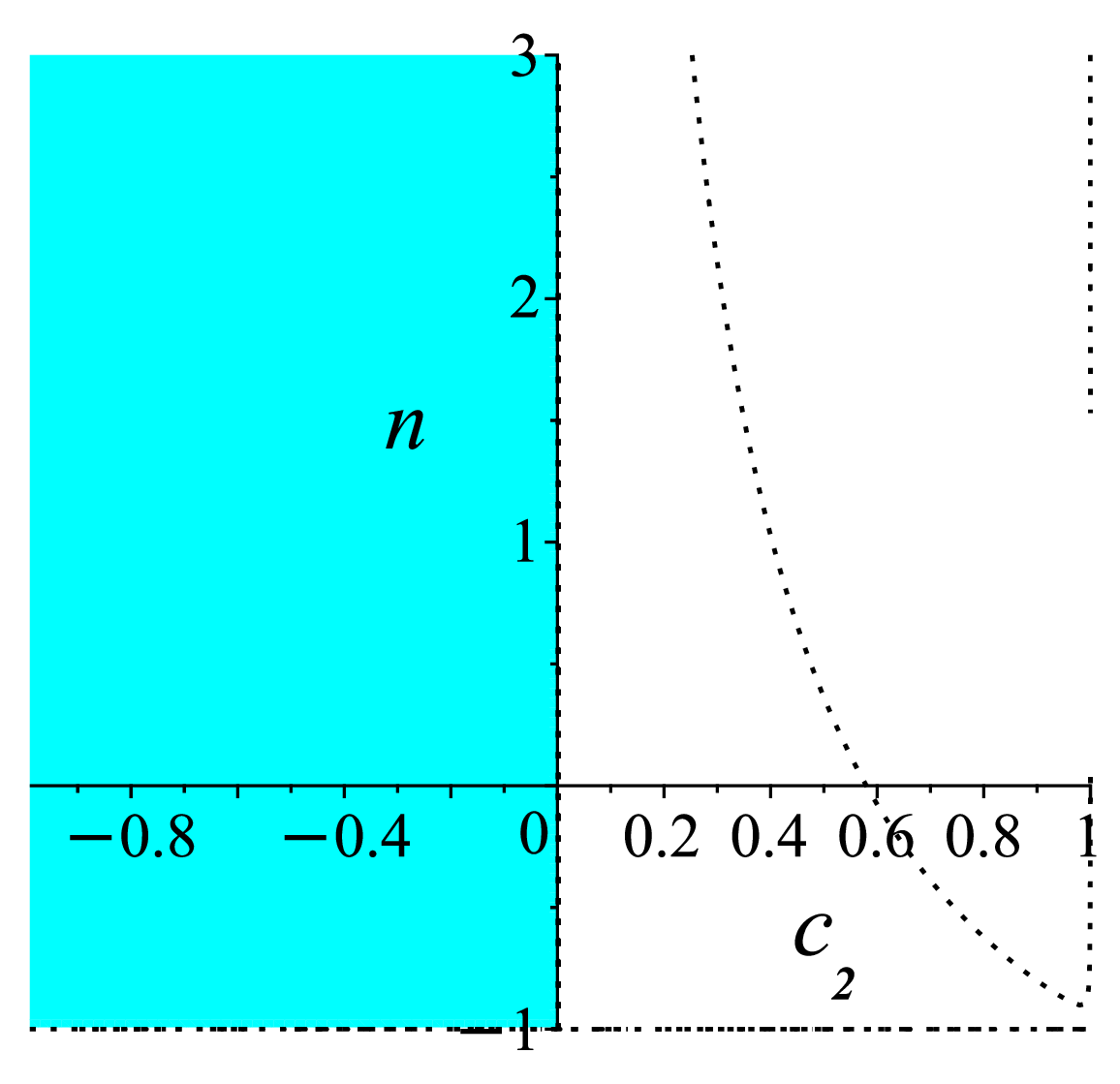}~~~~~~~ 	\includegraphics[height=6.3cm,width=7cm]{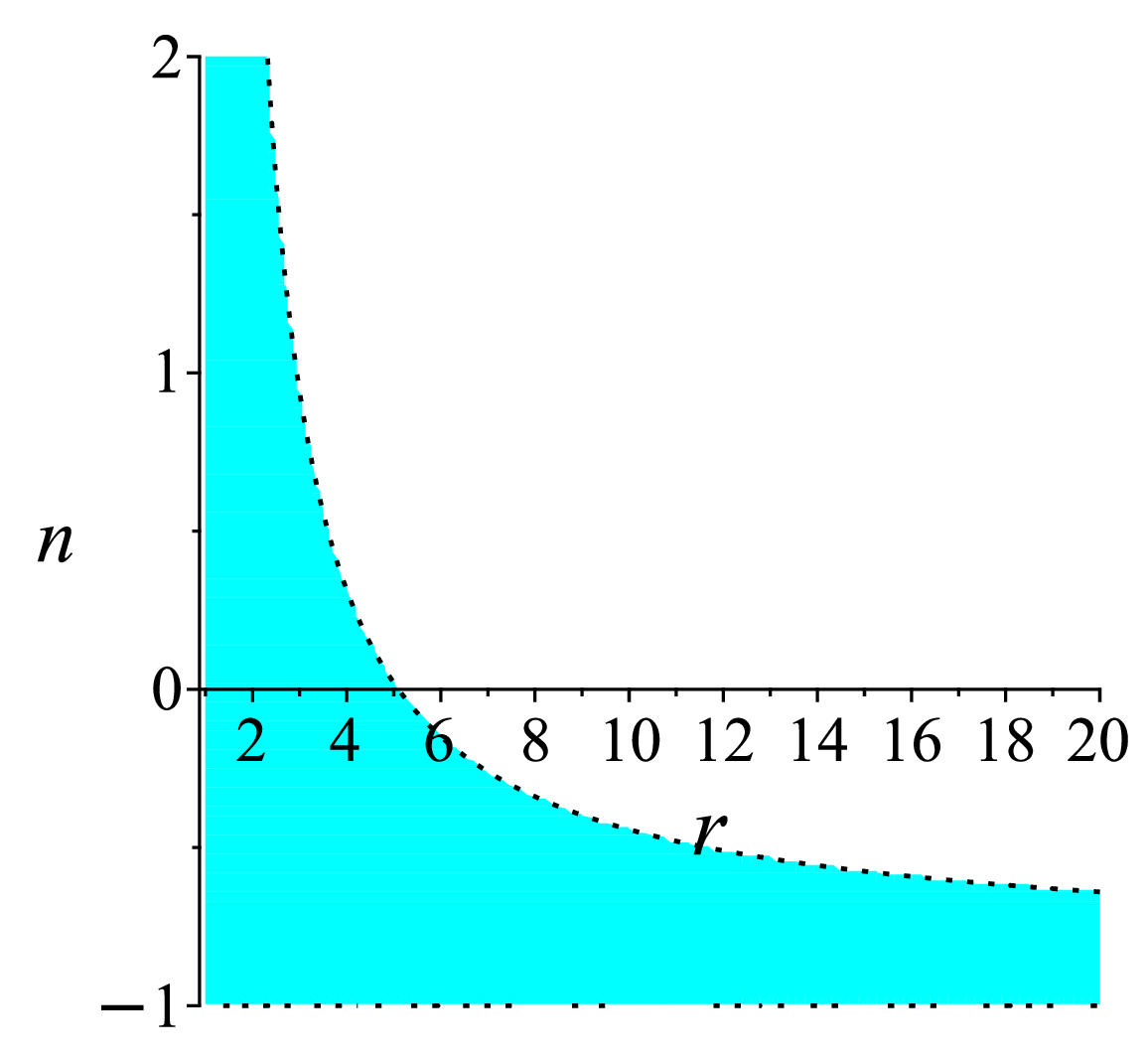}
		\caption{\small The effect of the exponent $ n $ on the energy conditions is illustrated in this figure. Here, we set $r_0=1$ and $j=1$. In panel (a) we set $r=1.02$ and it is evident that in the neighborhood of throat, WEC is respected for the range $-1< c_{2}<0$ regardless of the value of $n$. In (b) by setting $c_{2}=-1/2$, we find that satisfaction of  WEC occurs in more radial distances $r$ when $n$ decreases. } \label{nwn}
	\end{center}
\end{figure}

The role of exponent $n$ in satisfying the NEC and WEC, is explored in Fig. \ref{nwn} where the variation of $n$ together with $c_{2}$ (panel a) and $r$ (panel b) is considered to illustrate the regions in which energy conditions are satisfied. As shown in Fig. \ref{nwn}(a), where we set \(r=1.02\) to examine the region near the throat at \(r_0=1\), the WEC is satisfied throughout the allowed range \(-1< c_{2}<0\), independently of the value of the exponent \(n\). In panel (b), we fix \(c_{2}=-\frac{1}{2}\) and investigate the effect of \(n\) on the radial extent of the region in which the WEC is satisfied. We find that decreasing \(n\), within its allowed range \(n>-1\), extends this region farther away from the throat. We also examine the region in which \({\rm EC2}>0\) and \({\rm EC3}>0\), corresponding to the satisfaction of the NEC. The resulting behavior is qualitatively similar to that described above.

\subsection{The logarithmic shape function}
\begin{figure}
	\begin{picture}(0,0)(0,0)
		\put(100,-198){(a)}
		\put(340,-198){(b)}
	\end{picture}
	\begin{center}
		\includegraphics[height=6.3cm,width=7cm]{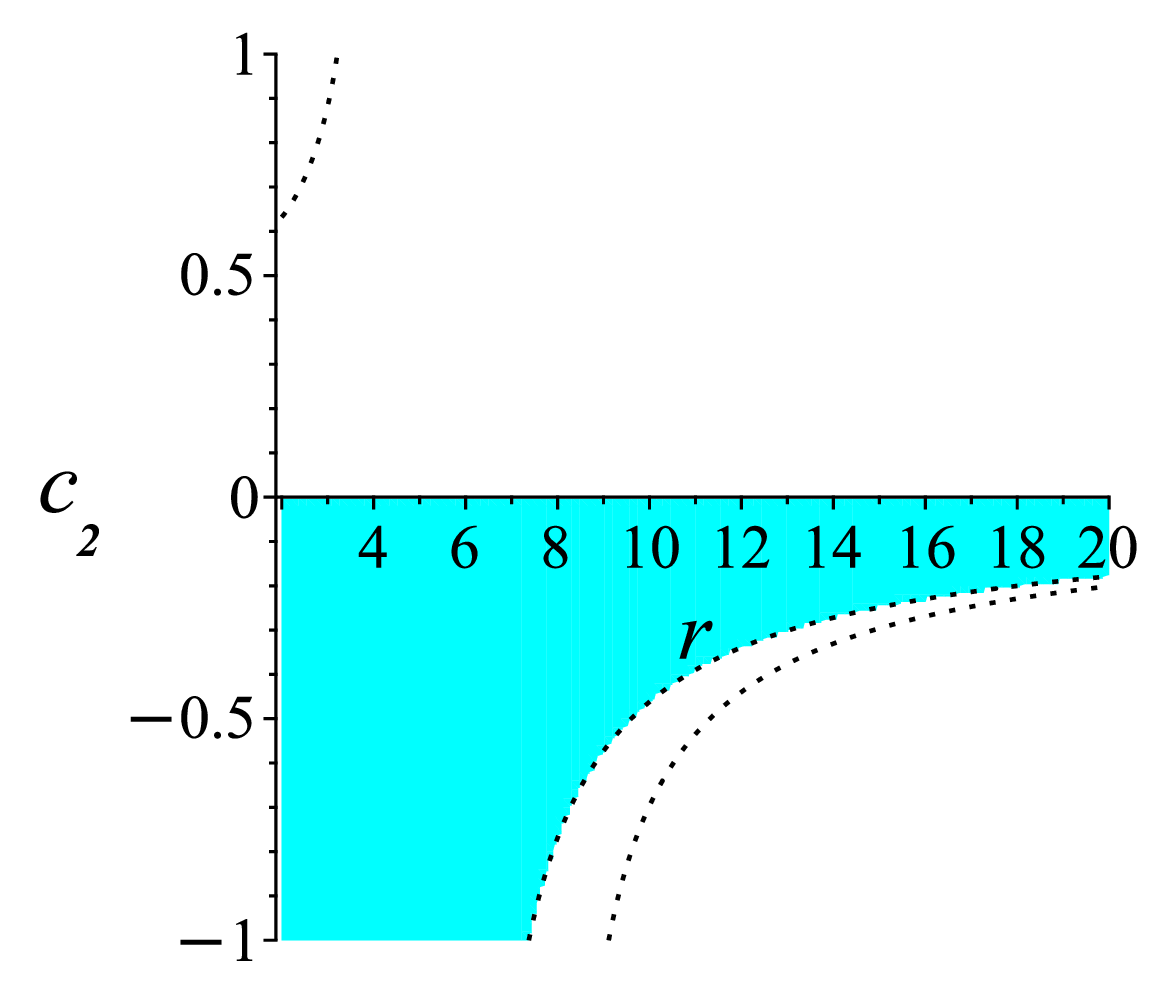}~~~~~~~~~~~~ 	\includegraphics[height=6.3cm,width=7cm]{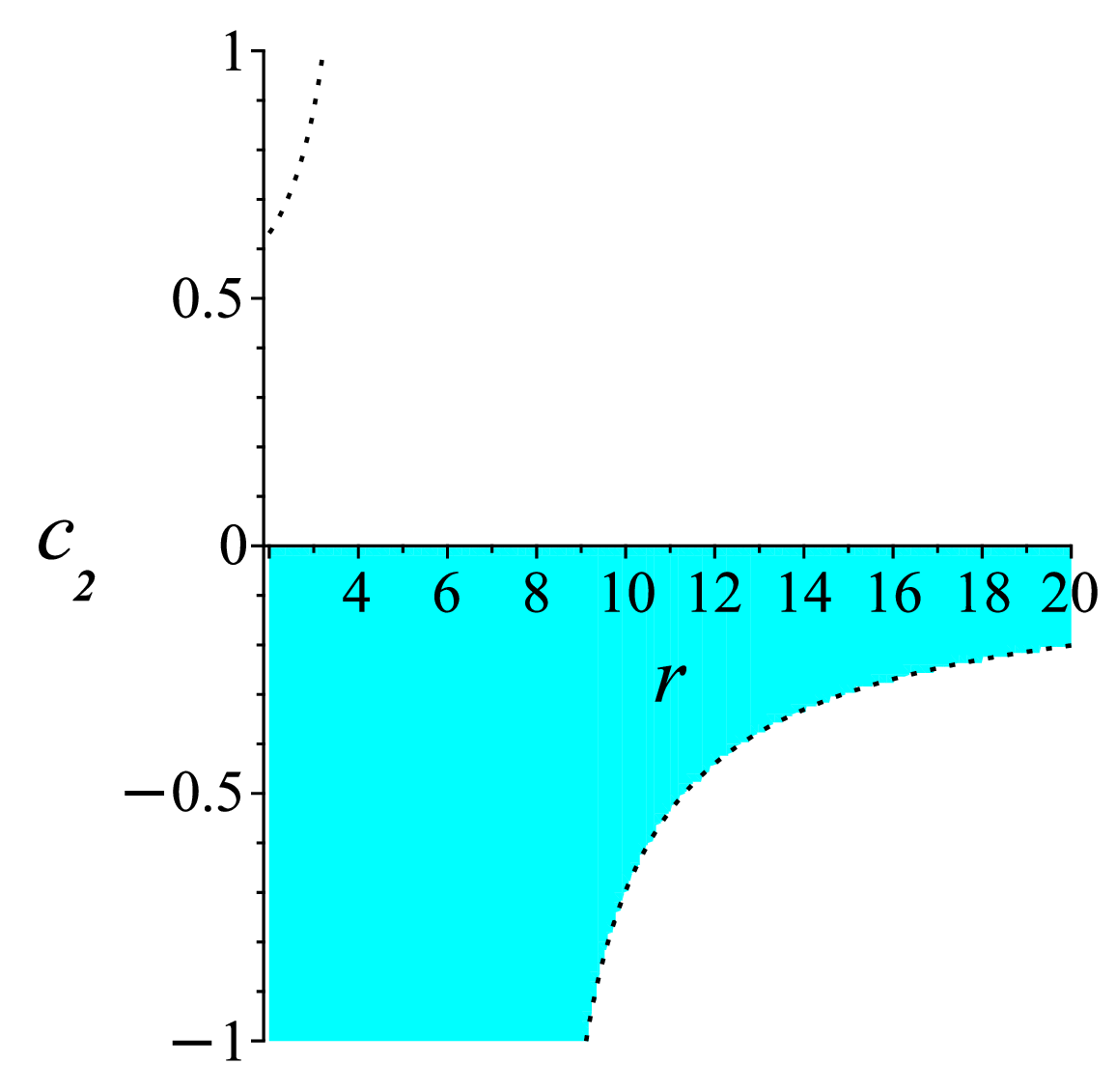}		
		\caption{\small In the blue region in (a) WEC and NEC are respected by the wormholes with logarithmic shape function. (b) shows the same region in which just NEC is respected. Here we set $r_0=2$ and $j=1$.} \label{lognecwec}
	\end{center}
\end{figure}

To find the energy conditions for wormhole solutions of the EA theory with logarithmic shape function, it is sufficient to substitute (\ref{logb}) into the energy density and pressures given by (\ref{rpp3}). Doing this, one finds
\bea
&&\begin{aligned}
\!\!\!\!{\rm EC1}&=\frac{1}{c_{2}r^4(\ln r)^2
		\left(\ln r-\ln r_0\right)^2}
	\bigg\{-c_{2}r^2(\ln r)^4
	+3c_{2}r^2\ln r_0(\ln r)^3
	\\
	&-\left[
	c_{2}^2j(r+j)\ln r_0
	-2jc_{2}r\ln r_0
	+r^2\ln r_0
	+3c_{2}r^2(\ln r_0)^2
	\right](\ln r)^2
	\\
	&+c_{2}r(\ln r_0)^2
	\left[r(\ln r_0+1)+jc_{2}\right]\ln r-c_{2}r^2(\ln r_0)^3
	\bigg\},
\end{aligned}\nn\\
&&\begin{aligned}
	\!\!\!\!{\rm EC2}=&\frac{\ln r_0}{c_{2}\,r^4(\ln r)^2
		(\ln r-\ln r_0)^2}
	\bigg\{ \left[
	jc_{2}(r+2j)
	-4jr	+\frac{2r^2}{c_{2}}
	\right](\ln r)^2
	\\
	&-r\ln r_0(r+jc_{2})\ln r -c_{2}r^2(\ln r_0)^2
	\bigg\},
\end{aligned}
\nn\\
&&\begin{aligned}
\!\!\!\!{\rm EC3}	&=\frac{1}{c_{2}\,r^4(\ln r)^2(\ln r-\ln r_0)^2}
	\Bigg\{ c_{2}r^2\left(\ln r-\frac12\right)(\ln r_0)^3
	\\
	&-\frac{c_{2}r}{4}
	\left[12r(\ln r)^2-7jc_{2}\ln r-jc_{2}+r\right](\ln r_0)^2
	\\
	& +c_{2}r
	\bigg[
	3r(\ln r)^3
	+\left(\frac{7}{4}j^2c_{2}^2
	-2jrc_{2}
	+\frac{r^2}{2}\right)(\ln r)^2
		\\
	&	+\left(-jc_{2}+r\right)r\ln r
	\bigg]\ln r_0
	-c_{2}r^2(\ln r)^4
	\Bigg\}.
\end{aligned}
\eea
In Fig. \ref{lognecwec}, we investigate the behavior of the energy conditions for wormholes with a logarithmic shape function. Panel (a) shows the region in which \({\rm EC1}>0\), \({\rm EC2}>0\), and \({\rm EC3}>0\) are simultaneously satisfied, corresponding to the fulfillment of both the WEC and NEC. Panel (b) illustrates the region where \({\rm EC2}>0\) and \({\rm EC3}>0\), indicating the satisfaction of the NEC. As can be seen, within the allowed range \(-1< c_{2}<0\), both the NEC and WEC are satisfied beyond the throat, located at \(r_0=2\). Moreover, for sufficiently small negative values of \(c_{2}\), these energy conditions are satisfied throughout the entire spacetime.

\subsection{The hyperbolic shape function}
\begin{figure}
	\begin{picture}(0,0)(0,0)
		\put(105,-219){(a)}
		\put(345,-219){(b)}
	\end{picture}
	\begin{center}
		\includegraphics[height=7cm,width=7cm]{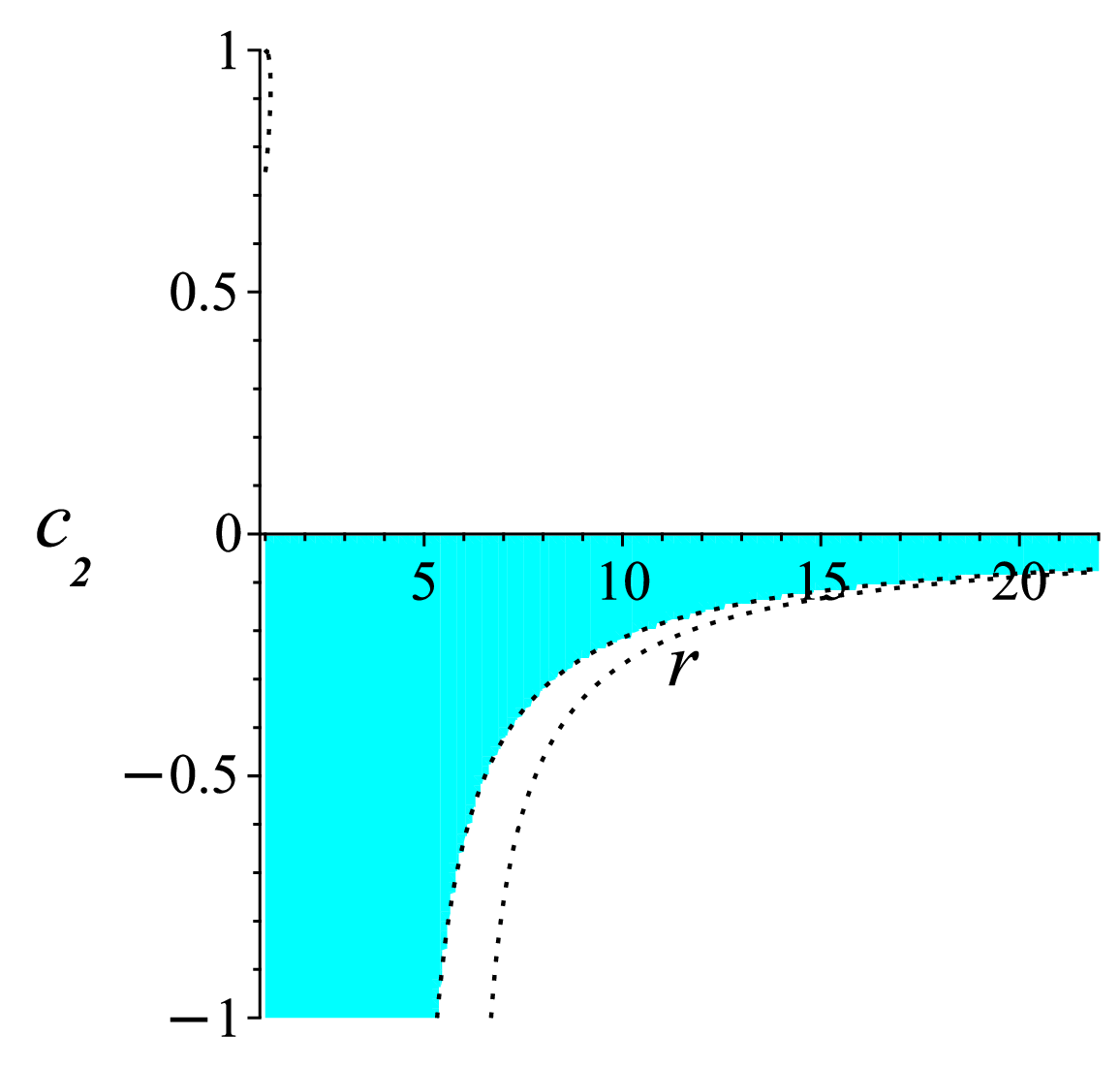}~~~~~~~~~~~~ 	\includegraphics[height=7cm,width=7cm]{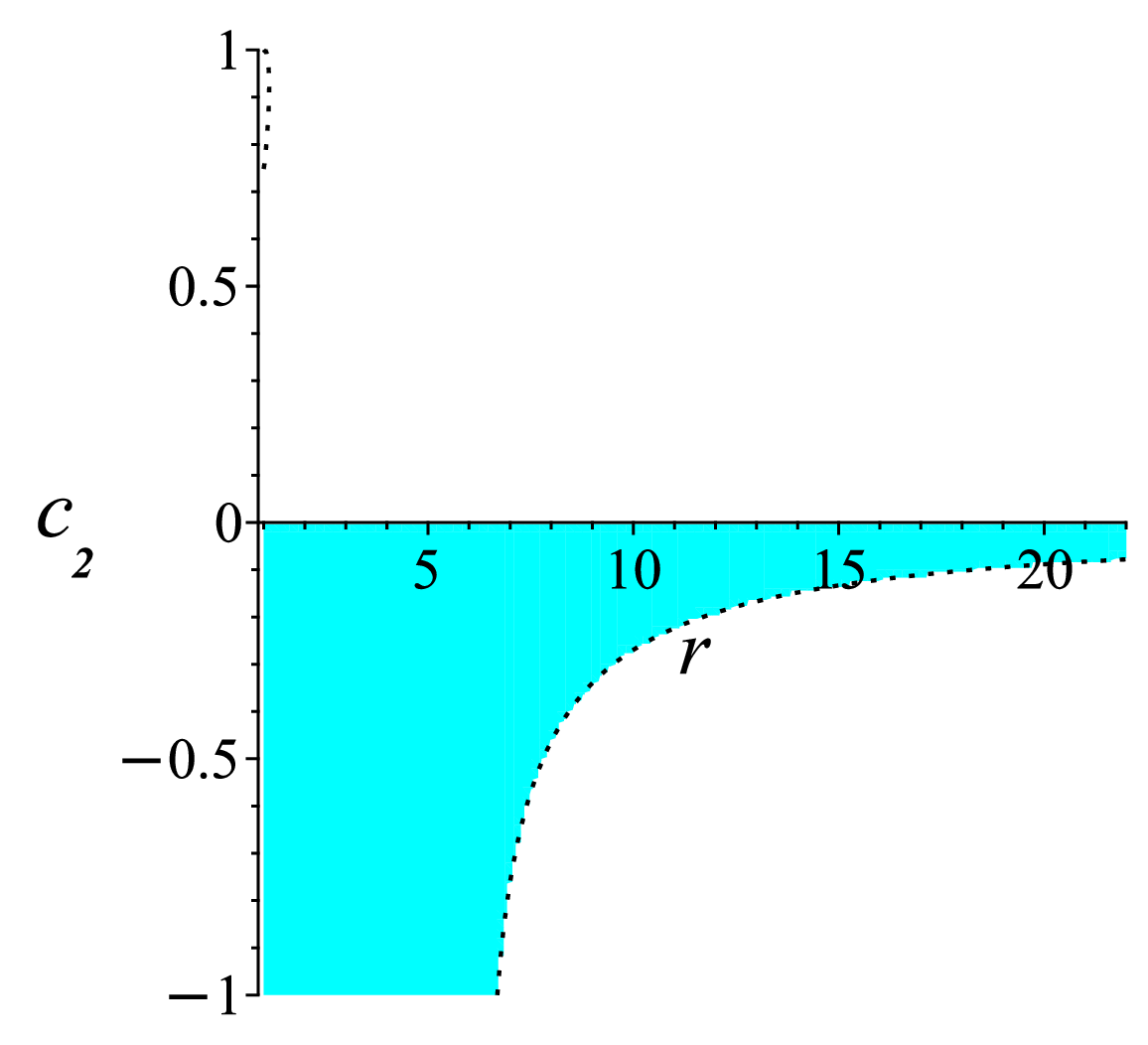}		
		\caption{\small Satisfaction of the energy conditions in the case of wormholes with hyperbolic shape function is depicted. In the blue region of (a) WEC and NEC are respected. (b) shows the same region just for NEC. Here, we set $r_0=1$ and $j=1$.} \label{hypnecwec}
	\end{center}
\end{figure}

For the wormhole solutions with the hyperbolic shape function in this class, the energy conditions can be obtained by inserting the hyperbolic shape function (\ref{hypb}) into the energy density and pressures (\ref{rpp3}), 
\bea
&&\begin{aligned}
\!\!\!{\rm EC1}&=\frac{1}{c_{2}\,r^3
	\left(rS_0C-r_0SC_0\right)^2}\bigg\{\!-\left[(2r-Jc_{2})S^2\!
	+r\frac{S}{C}\Big(r-r_0\frac{SC_0}{S_0C}+Jc_{2}\Big)
	\right]\!r_0^2c_{2}C_0^2
	\\
	&+S_0r_0\Big[
	-\left(
	\left(1-3c_{2}\right)r^2
	+Jc_{2}(c_{2}-2)r
	+c_{2}^2J^2
	\right)CS
	\\
	&
	+\left(
	r^2+Jc_{2}(c_{2}-2)r+c_{2}^2J^2
	\right)r
	\Big]C_0-c_{2}r^3S_0^2C^2\bigg\},
\end{aligned}\nn\\
&&\begin{aligned}
\!\!\!{\rm EC3}&=\frac{2r_0(SC-r)C_0}{c_{2}r^3(rS_0C-r_0SC_0)^2S_0C_0^2}\bigg\{2r_0\Big[
S_0^2
\Big(
J(r+2J)c_{2}^2
-4Jc_{2}r+2r^2
\Big)C^2
\\
&-S C_0 S_0 c_{2}r_0
	(Jc_{2}+r)C
+S^2C_0^2c_{2}r_0^2
\Big]\bigg\},
\end{aligned}\nn\\
&&\begin{aligned}
\!\!\!{\rm EC3}	&=\frac{1}{4c_{2}r^3	\left(rS_0C-r_0SC_0\right)^2}\bigg\{-r_0^2c_{2}C_0^2
	\bigg[
	\Big(-2r_0\frac{SC_0}{CS_0}
	-6Jc_{2}+11r\Big)S^2
	\\
	&+
	r\bigg(
	-2r_0\frac{C_0S^2}{S_0C^2}
	+\left(3Jc_{2}+4r\right)\frac{S}{C}
	+r\Big(\frac{1}{C^2}-4\Big)
	(r-Jc_{2})
	\bigg)
	\bigg]
	\\
	&+r_0S_0C_0	\bigg[
	-\left(J\left(9r+4J\right)c_{2}^2
	-8r\left(2r+J\right)c_{2}
	+4r^2
	\right)CS
	\\
	&	+r\Big(4(Jc_{2}-r)r^2c_{2}\frac{S}{C}
	+\left(9rJ+4J^2\right)c_{2}^2
	-r(r+2J)c_{2}+r^2
	\Big)
	\bigg]-4c_{2}r^3S_0^2C^2\bigg\}.
\end{aligned}
\eea
The same analysis shows that, these wormhole solutions respect the NEC and WEC beyond the throat $r_0$ and also {\it throughout the whole space} outside the $r_0$, if the coupling constants of the EA theory chosen adequately in the range $-1< c_{2}<0$. Fig. \ref{hypnecwec} shows the parameter space of these wormhole solutions. Similar to the previous cases, in order to satisfy the NEC and WEC faraway from the throat, the coupling combination $c_{2}$ should take small negative values.

We conclude this section by summarizing our main findings. Wormholes with power-law, logarithmic, and hyperbolic shape functions can exist in Einstein-aether (EA) theory with the coupling constants satisfying \(c_2=-c_{13}\neq 0\) and \(c_{14}=0\). A distinctive feature of this class is that, for sufficiently small negative values of the coupling \(c_{2}\), the energy conditions can be satisfied throughout the entire spacetime.
	
Our results further demonstrate that the fulfillment of NEC and WEC for these wormhole configurations requires negative values of the aether coupling constant \(c_2\). Accordingly, the condition \(c_2<0\) can be regarded as a necessary requirement for the ``absence of exotic matter" in Class III. This condition can be combined with the constraints given in (\ref{ccon}), which, for the present class of solutions, yield the lower bound \(c_2>-1\). Combining these two requirements, we conclude that the coupling constant \(c_2\) must lie within the range
\be	\label{conexo2}
	-1<c_2<0.
\ee
It should also be noted that an additional constraint arises from the propagation of gravitational waves in Einstein–Aether theory \cite{Gong:2018cgj}, which requires $\frac{c_{14}}{1-2c_{14}}<c_2$.	Since \(c_{14}=0\) for the present class, this constraint reduces to \(c_2>0\). Consequently, as in the case of Class I, the condition for the absence of exotic matter, given by (\ref{conexo2}), is excluded. This implies that, as in the case of Einstein gravity, exotic matter is necessary to support the wormhole geometries in this class of Einstein–Aether theory. 


\section{Summary}\label{summ}\label{sec7}

In the context of Einstein gravity, wormhole geometries typically violate the energy conditions due to the flare–out condition at the wormhole throat \cite{Hochberg:1997wp}.
In order to investigate wormhole geometries that can be supported by ordinary (non-exotic) matter, we considered traversable wormholes within the framework of Einstein-aether theory. By adopting an anisotropic energy–momentum tensor, we solved the equations of motion of the EA theory for specific combinations of the coupling constants. We then derived the corresponding energy density and pressure components for the wormholes belonging to the three identified classes in the EA theory, as follows:
\begin{itemize}
	\item {\rm Class I:} ~~~ $c_2\neq 0$, $c_{13}=0$, and $c_{14}=0\,,$
	\item {\rm Class II:} ~~ $c_2=0$, and $c_{14}=0\,,$
	\item {\rm Class III:} ~ $c_2=-c_{13}\neq0$, and $c_{14}=0\,$.
\end{itemize}
From previous observational and theoretical studies, several constraints on the coupling constants $ c_i $ of the EA theory have been obtained, as follows \cite{Ding:2015kba,Jacobson:2007fh,Yagi:2013ava}
\be \label{ccon2}
0\le c_{14}<2\,, \qquad 2+c_{13}+3c_2>0\,, \qquad 0\le c_{13}<1\,.
\ee
In this work, we investigate the NEC and WEC for wormhole geometries by considering three different types of shape functions, as follows
\begin{itemize}
	\item {\rm Power law shape function:} ~~~ $b(r)=r_0 \left(r_0/r \right)^n\,,$
	\item {\rm Logarithmic shape function:} ~~~ $b(r)=r \lp \ln r_0/ \ln r \rp\,,$
	\item {\rm Hyperbolic shape function:} ~~~ $b(r)=r_0 \lp \tanh r/ \tanh r_0 \rp\,,$
\end{itemize}
for each classes of coupling values. 

Our results show that wormhole geometries can satisfy the NEC and WEC both in the vicinity of the throat and beyond it, provided that the parameters \(r_0\), \(n\), and \(c_i\) are appropriately chosen.
	
In Class I, our analysis shows that the NEC and WEC can be satisfied when the aether coupling \(c_2\) is negative. Furthermore, the constraints given in Eq. (\ref{ccon2}) impose the lower bound \(c_2\geq -\frac{2}{3}\). Combining these two conditions, we obtain the allowed range
	\be \label{coex1}
	-\frac{2}{3}\leq c_2<0,
	\ee
	which can be regarded as the condition for the absence of exotic matter in this class.
	
Our results for Class II show that the NEC and WEC can be satisfied only when the coupling combination $c_{13}$ takes negative values. However, this requirement is in contradiction with the constraints given in Eq. (\ref{ccon2}). Therefore, for physically allowed values of $c_{13}$, wormhole geometries belonging to this class necessarily violate the energy conditions as in Einstein gravity.
	
	In Class III, we also find that the wormhole solutions can satisfy the energy conditions for negative values of the coupling constant \(c_2\). Thus, the condition \(c_2<0\) can be regarded as a requirement for the absence of exotic matter in this class. Furthermore, the constraint given in Eq. (\ref{ccon2}) yields the lower bound \(c_2>-1\). Combining these two conditions, we obtain the following allowed range of \(c_2\) for wormhole solutions satisfying the energy conditions:
	\be
	\label{coex2}
	-1<c_2<0.
	\ee

However, an additional constraint on the coupling constants was derived in \cite{Gong:2018cgj}, namely, $\frac{c_{14}}{1-2c_{14}}<c_2$. Since \(c_{14}=0\) for  Classes I and III, this constraint reduces to
\be
c_2>0.
\ee
This bound is incompatible with the parameter ranges obtained in Eqs. (\ref{coex1}) and (\ref{coex2}). Consequently, the NEC and WEC cannot be satisfied by the wormhole geometries in Classes I and III within the physically allowed parameter space. This suggests that, as in the Einstein gravity, exotic matter is necessary to support traversable wormholes in the EA theory.
	
\vspace{2mm}

\appendix

\textbf{Conflict of Interest:}  The authors declare that they have no conflict of interest.

\textbf{Data availability statement:} The authors declare that the data supporting the findings of this study are available within the paper. The results presented in this paper are derived through analytical methods and therefore, do not rely on any empirical data. 

\providecommand{\href}[2]{#2}\begingroup\raggedright
\endgroup
\end{document}